\documentclass[twocolumn]{aastex62}
\usepackage{natbib}
\usepackage{bm}
\usepackage{graphicx}
\usepackage{physics}
\usepackage{mathtools}
\usepackage{url}
\received{January 1, 2018}
\revised{January 7, 2018}
\accepted{\today}
\submitjournal{ApJ}

\shorttitle{Non-thermal emissions from a head-tail galaxy}
\shortauthors{Ohmura et al.}
\begin{document}

\title{Non-thermal emissions from a head-tail radio galaxy in 3D magnetohydrodynamic simulations}

\correspondingauthor{Takumi Ohmura}
\email{tohmura@icrr.u-tokyo.ac.jp}

\author[0000-0002-0040-8968]{Takumi Ohmura}
\affil{Institute for Cosmic Ray Research, The University of Tokyo, \\
5-1-5 Kashiwanoha, Kashiwa, \\ 
Chiba 277-8582 Japan}

\author[0000-0001-9064-160X]{Katsuaki Asano}
\affil{Institute for Cosmic Ray Research, The University of Tokyo, \\
5-1-5 Kashiwanoha, Kashiwa, \\ 
Chiba 277-8582 Japan}

\author[0000-0003-2370-0475]{Kosuke Nishiwaki}
\affil{Institute for Cosmic Ray Research, The University of Tokyo, \\
5-1-5 Kashiwanoha, Kashiwa, \\ 
Chiba 277-8582 Japan}

\author{Mami Machida}
\affiliation{Division of Science, National Astronomical Observatory of Japan, \\ 
2-21-1 Osawa, Mitaka, \\
Tokyo 181-0015, Japan}

\author[0000-0002-4037-1346]{Haruka Sakemi}
\affiliation{Graduate School of Science and Engineering, Kagoshima University, \\
1-21-35 Korimoto, \\
Kagoshima 890-0065, Japan}

\begin{abstract}
We present magnetohydrodynamic simulations of a jet-wind interaction in a galaxy cluster and the radio to gamma-ray and the neutrino emissions from this "head-tail galaxy". 
Our simulation follows the evolution of cosmic-ray (CR) particle spectra with energy losses and the stochastic turbulence acceleration.
We find that the reacceleration is essential to explain the observed radio properties of head-tail galaxies, in which the radio flux and spectral index do not drastically change.
Our models suggest that hard X-ray emissions can be detected around the head-tail galaxy in the Perseus cluster by the hard X-ray satellites, such as FORCE, and it will potentially constrain the acceleration efficiency.
We also explore the origin of the collimated synchrotron threads, which are found in some head-tail galaxies by recent high-quality radio observations.
Thin and elongated flux tubes, connecting the two tails, are formed by strong backflows at an early phase.
We find that these threads advect with the wind for over 300 Myr without disrupting.
The radio flux from the flux tubes is much lower than the typical observed flux.
An efficient CR diffusion process along the flux tubes, however, may solve this discrepancy.

\end{abstract}


\section{Introduction} \label{sec:intro}
% WATについて簡単に
Radio jets from active galactic nuclei (AGN) are observed in clusters of galaxies.
% Some radio jets, which is called 'head-tail' radio galaxies, have 'U' \kn{or} 'V' shape.
Radio jets with 'U' or 'V' shape are often termed as 'head-tail' galaxies.
These sources are composed of a luminous AGN ('head') and diffuse radio plumes of bent jets ('tail') \citep{1968MNRAS.138....1R,1972Natur.237..269M}.
Tails (or 'lobe' for standard radio jets) are spatially extended on scales of hundreds of kpc, in which a large number of relativistic cosmic-ray particles are stored.
% Therefore, radio jets are generally one of the ideal sites to study high-energy astronomical phenomena.
The standard scenario for the formation of head-tail galaxies is that the curved tails are formed due to the ram pressure induced by the peculiar motion of the host galaxy in the intracluster medium (ICM) and/or the large-scale turbulence motion of ICM \citep{1979Natur.279..770B,1979ApJ...234..818J}.
Another scenario is that strong magnetic fields, which are amplified by the motion of ICM, bend the radio jets by magnetic tension force \citep{1997ApJ...488..572S,2021Natur.593...47C}.
The models of the ram pressure bending are well supported by several hydrodynamics and magnetohydrodynamics (MHD) simulations \citep{1984Natur.310...33W,1992ApJ...393..631B,2017ApJ...839...14G,2019ApJ...884...12O}.
% In particular, MHD studies successfully explain the morphology of head-tail galaxy.

% Synchrotron Threadsについて
Another interesting finding in recent radio observations of head-tail galaxies is collimated synchrotron threads, whose width is a few kpc and length is several tens kpc \citep{2020A&A...636L...1R,2021Natur.593...47C,2022A&A...657A..56K,2022ApJ...935..168R}.
In particular, MeerKAT observations have firstly shown that a head-tail galaxy ESO 137-006 has threads liking the tails with a steep spectral index of about 2 between 1000 MHz and 1400 MHz \citep{2020A&A...636L...1R}.
However, the origin of these threads is not well understood.
These threads provide us useful insights into the physical process of CRe transport and the coherent scale of the magnetic field.
Several MHD simulations of galaxy cluster mergers with AGN jets show that the threads-like structures can be formed by merger-driven flows that stretch out the old CRe from AGN jets \citep{2021A&A...653A..23V,2021ApJ...914...73Z}.

% WATの再加速の理論モデル
Reacceleration of cosmic-ray protons (CRp) and electrons (CRe) due to the Fermi-II type stochastic process \citep{PhysRev.75.1169} is often invoked to explain the observed morphology of head-tail galaxies.
%is an energy gain of cosmic-ray protons (CRp) and electrons (CRe) due to the Fermi-II type stochastic process \citep{PhysRev.75.1169}.
%Head-tail galaxies are one of the ideal celestial laboratories to study this process.
The origin of the diffuse radio emission of the tails is synchrotron radiation from CRe.
Several radio observations reveal that the tailed region retains a roughly constant radio flux and spectral index with the spatial extent of up to 
several hundreds kpc \citep{1975A&A....38..381M,1986ApJ...301..841O,1998A&A...331..475F}.
The timescale of the energy loss of CRe in a cluster magnetic field \citep[$\sim \mu$G; ][]{2004IJMPD..13.1549G} is shorter than the times-scale of wind advection.  
Therefore, the reacceleration process of CRe is needed to explain radio properties \citep{1976ApJ...203..313P}.
Since no radio and X-ray shocks are observed at tails, the turbulence driven by the jet-ICM interaction is thought to play a major role in the acceleration of CRe \citep[e.g.,][]{1989ApJ...336..243S,1988SvAL...14..255P}.
Recently, gentle reacceleration with a timescale longer than 100 Myr has been suggested to explain the radio features of the head-tail galaxies in Abell 1033 \citep{2017SciA....3E1634D, 2022A&A...666A...3E} and 2A0335+096 \citep{2022A&A...659A..20I}. 
The turbulent reacceleration would also play an important role in various diffuse sources such as the radio halos in galaxy clusters \citep{2001ApJ...557..560P, 2014IJMPD..2330007B,2015ApJ...815..116F,2022ApJ...934..182N}, the Fermi bubbles \citep{2011PhRvL.107i1101M, 2015ApJ...814...93S}, and pulsar wind nebulae \citep{2017ApJ...841...78T}.
% , gamma-ray burst \citep{2009ApJ...705.1714A, 2016PhRvD..94b3005A}, and blazers \citep{2014ApJ...780...64A}.

% シミュレーションによるCReの取り扱い
To study the connection between the jet dynamics and non-thermal processes, fluid simulations combined with CR evolution are essential.
% Although the first principle approach, such as particle-in-cell simulations or hybrid simulations, can treat the particle acceleration process naturally, it can not handle fluid scale due to high computational costs.
One major approach is to solve Fokker--Planck equations in the MHD simulations simultaneously.
% So that CR particles are advected passively with respect to bulk flow.
This phenomenological approach is valid when the gyration radius of CR particles is much less than MHD grid scales and CR particles have nearly an isotropic equilibrium distribution function as a consequence of frequent pitch-angle scattering on sub-grid scale MHD turbulence. 
Nowadays, several groups have successfully developed simulation codes with this approach \citep{2005APh....24...75J,2009ApJ...696.1142M,2018ApJ...865..144V,2019MNRAS.488.2235W,2021A&A...653A..23V}.
% 再加速を取り入れた先行研究
Several groups have demonstrated the role of reacceleration for the non-thermal emissions for galaxy clusters \citep{2013ApJ...762...78Z,2014MNRAS.443.3564D} and powerful AGN jets \citep{2021ApJ...921...74K,2022A&A...667A.138K}. 
% From a macroscopic viewpoint, the acceleration due to the Fermi II order mechanism is regarded as the momentum diffusion in phase space. 
% The momentum diffusion coefficient $D_{\gamma \gamma}(\gamma)$, where $\gamma$ is the Lorentz factor of CR particles, depends on the micro-physics and property of MHD turbulence \citep[e.g.,][]{1989ApJ...336..243S,1988SvAL...14..255P,2007MNRAS.378..245B}.
% Commonly, the form of $D_{\gamma \gamma}$ in MHD simulation is given as a function of the fluid variables in a 'subgrid' way.

% 我々のアプローチについて
In this paper, we explore multi-wavelength emissions from radio to gamma-ray and neutrino emissions from a head-tail galaxy in our MHD simulation.
Our code follows the energy and spatial evolutions for CRe and CRp, which are accelerated by sub-grid scale turbulence.
To connect the strength of turbulence and the acceleration efficiency, we employ a sub-grid model. 
% which is detailed in section \ref{subsec:reacc.}.
% After the simulations, we analyzed the emission features, calculated in the post-processing way.
Our aim is on constraining the acceleration efficiency.
We also investigate the origin of the collimated synchrotron threads in a simulated head-tail galaxy.

% ロードマップ
Our paper is structured as follows: in section \ref{sec:setup} we present the setup of our MHD simulations and numerical methods for solving the Fokker--Planck equations employed in this paper, and we propose our 'sub-grid' model for reacceleration in section \ref{subsec:reacc.}.
In section \ref{sec:Results} and \ref{sec:emission}, we present the results of the MHD simulations and the murti-wavelength and the neutrino emissions of the simulated head-tail galaxy.
We discuss the transport mechanism of CRe to reproduce the collimated synchrotron threads in section \ref{sec:diffusion}.
Section \ref{sec:summary} summarizes our results and discusses future developments of this work.

\section{Numerical Method} \label{sec:setup}
\subsection{Simulation setup}
Our simulation tracks the evolution of AGN jets extending up to hundred kpc scales in hundred Myrs \footnote{Movies of our simulation are available in \url{https://www.youtube.com/playlist?list=PLgnUM4yGp9oKG4ZQrsJC6oVbvDu7n3JFp}}.
To simulate the dynamics of a head-tail galaxy, we follow the simulation setup by \citet{2019ApJ...884...12O}.
The jet would be launched with a highly relativistic speed from the AGN, but it would be strongly decelerated on the kpc scale \citep[e.g.,][]{1984ApJ...286...68B}.
Thus, we deal with non-relativistic jets and solve ideal MHD equations.
The MHD simulations are carried out by using deeply modified version of CANS+ \citep{2019PASJ...71...83M}.
CANS+ employs the HLLD Riemann solver to compute the numerical flux across grid interfaces \citep{2005JCoPh.208..315M}.
The time integral is performed with the third-order strong-stability-preserving  Runge-Kutta method, and reconstruction adopts a fifth-order-monotonicity-preserving interpolation scheme \citep{1997JCoPh.136...83S}.
The hyperbolic divergence cleaning method is adopted to maintain the condition of $\nabla \cdot \bm{B}=0$ \citep{2002JCoPh.175..645D}.

We use a Cartesian coordinate $(x,y,z)$ with a uniform cells, $\Delta = 0.375$ kpc, and the numerical resolution is chosen to be $N_{x} \times N_{y} \times N_{z} = 640 \times 880 \times 280$.
Thus, the simulation domain is defined by $x \in [-20, 220]$ kpc, $y \in [-164, 164]$ kpc, and $z \in [-52, 52]$ kpc. 
The jets are launched along $\pm y$-directions.
We impose free boundary conditions in the $+x$, $\pm y$, and $\pm z$ directions.
The gas adiabatic index is simply constant as $\Gamma = 5/3$.
In this simulation, we solve simultaneously the Fokker--Planck equations for CRe and CRp including particle reacceleration process with MHD equations (see detail in the next sub-section).
We perform simulations for three models with varying acceleration efficiencies.
The model without reacceleration only considers the evolution of the CRe spectra
The symmetrical boundary condition is applied at $y = 0$ for only the models with reacceleration for reducing the computational cost.

The bipolar jets propagate in unmagnetized and uniform ICM, whose number density and temperature are $n_{\rm ICM} = 5.0 \times 10^{-3}$ ${\rm cm^{-3}}$ and $T = 5$ keV, respectively.
To implement ICM winds, we set initial flow velocity along x-direction for the ICM and an incoming flow boundary condition on the -x boundary.
The wind velocity, $v_{\rm ICM}$, is $400$ ${\rm km/s}$.
The central black hole is located at the coordinate origin, and the jet material is injected into the simulation domain through an area of circle with a radius $r_{\rm jet}=$ 3 kpc at a distance 6 kpc from the origin.
The density, sonic Mach number, and temperature for jets are $0.01 n_{\rm ICM}$, 2, and 500 keV, respectively.
The jets are initially weakly magnetized with a purely toroidal magnetic field $B_{\phi} = B_{\rm jet} {\rm sgn}(z) \sin^4 (\pi r/r_{\rm jet}),$ where $B_{\rm jet} = 6.8$ $\mu$G \citep{2014ApJ...789...79A}.
This simple configuration of magnetic field is valid only for the high-beta jets ($\beta_{\rm jet} \equiv 8\pi p_{\rm jet}/B_{\rm jet}^2 = 25$).
Hence, our jets are a kinetic energy dominant with the kinetic power $L_{\rm kin} = 2.0 \times 10^{44}$ $\rm erg~s^{-1}$.
To make non-axisymmetric features, a small-amplitude (1 percent) random pressure perturbation of gas pressure for the jet flows is adapted at each grids \citep{2019MNRAS.490.4271M}.
On direct cross-wind and non-relativistic jet interaction, the balance between the ram pressures provides a characteristic bending length, $l_{\rm b} \approx r_{\rm jet} (\rho_{\rm jet}v_{\rm jet}^2)/(\rho_{\rm ICM}v_{\rm ICM}^2) \sim 70$ kpc \citep{1979ApJ...234..818J}.
The parameters of the jets and ICM are summarized in Table \ref{tab:model}.

\subsection{The evolution of CRe and CRp}
In our simulations, CR particles injected in the jet flows are advected passively with the background MHD bulk flow, with the bulk velocity denoted by $\bm{v}$. 
This simplification implies strong coupling between the thermal plasma and CR particles due to the wave-particle interaction.
As mentioned before, we do not identify the CR pressure independently in our MHD simulations, i.e., the two-fluid approximation is not adopted.
While the CR injection after the jet injection is ignored throughout this simulation, CR particles are accelerated by sub-grid scale turbulence.
The reacceleration process is treated phenomenologically as energy diffusion process, and we introduce a sub-grid model to determine the acceleration efficiency.

We solve the Fokker--Planck equations of CRe and CRp without injection term \citep[e.g.,][]{2002cra..book.....S}:
\footnotesize
\begin{eqnarray}
    \label{eq:CRe1}
    \pdv{N_e}{t} + \nabla \cdot (N_e \bm{v}) &=&  
     \pdv{\gamma}\qty[N_e \frac{\gamma}{3}(\nabla \cdot \bm{v}) - N_e \dot{\gamma}_{\rm cool}] \nonumber \\ 
    &-& \pdv{\gamma} \qty[\frac{N_e}{\gamma^2} \pdv{\gamma}\qty(\gamma^2 D_{\gamma \gamma})]+\pdv[2]{\gamma}\qty[D_{\gamma\gamma}N_e],
\end{eqnarray}
\begin{eqnarray}
    \label{eq:CRp1}
    \pdv{N_p}{t} + \nabla \cdot (N_p \bm{v}) &=& \pdv{\gamma}\qty[N_p \frac{\gamma}{3}(\nabla \cdot \bm{v}) ] \nonumber \\ 
    &-& \pdv{\gamma} \qty[\frac{N_p}{\gamma^2} \pdv{\gamma}\qty(\gamma^2 D_{\gamma \gamma})]+\pdv[2]{\gamma}\qty[D_{\gamma\gamma}N_p],
\end{eqnarray}
\normalsize
where $N_{\rm e,p}(\gamma, \bm{x}, t)$ are the number densities of CRe and CRp with Lorentz factor $\gamma$,
respectively, and $\dot{\gamma}_{\rm cool}$ is the energy loss function.
The second line for both the equations represents the energy diffusion by the stochastic reacceleration process (Fermi-II reacceleration) with the diffusion coefficient $D_{\gamma \gamma}(\gamma)$.
For simplicity, we ignore the back-reaction of CR particles to the fluid, the loss process by $pp$-collision and the Coulomb collision of CRp, and the spatial diffusion for both CRe and CRp.

For the energy loss of CRe, we consider the Coulomb collision, synchrotron radiation, and inverse Compton scattering with cosmic microwave background (CMB) photons, so that $\dot{\gamma}_{\rm cool} = \dot{\gamma}_{\rm C} + \dot{\gamma}_{\rm rad} + \dot{\gamma}_{\rm IC}$.
The energy loss rate by the Coulomb collision is given by \citep{GOULD1972145,2019MNRAS.488.2235W}
\begin{eqnarray}
    \dot{\gamma}_{\rm C} = - \frac{3\sigma_{\rm T} n_{\rm e} c}{2} \left\{ \ln \qty( \frac{m_{\rm e}c^2 
    \sqrt{\gamma-1} }{\hbar \omega_{\rm pl} } ) 
        + \ln(2) \qty(\frac{1}{2}+\gamma^{-1}) \nonumber \right. \\  \left.
    + \frac{1}{2} + \left(\frac{\gamma-1}{4\gamma} \right)^2 \right\} ,
\end{eqnarray}
where $n_{\rm e}$ is the number density of thermal electrons, and $\omega_{\rm pl}$ is the electron plasma frequency.
The loss rates by the synchrotron radiation and inverse Compton scattering are
\begin{equation}
    \dot{\gamma}_{\rm rad} = - \frac{4\sigma_T}{3m_{\rm e} c^2} \gamma^2 \frac{B^2}{8\pi},
\end{equation}
\begin{equation}
    \dot{\gamma}_{\rm IC} = - \frac{4\sigma_T}{3m_{\rm e} c^2} \gamma^2 u_{\rm CMB},
\end{equation}
where $B$ and $u_{\rm CMB}(z)$ are the magnetic field and the CMB photon energy density at redshift $z$, respectively.
This paper adopts a constant redshift, $z = 0$, for simplicity.
We assume that the magnetic field is disturbed on much smaller scales than the numerical grid, so that the pitch angle distribution for CR particles is isotropic.
%In our simulations, the AGN is located at $z=0$ for simplicity. 
% This paper adopt $z = 0$.

To solve the Fokker--Planck equations of CRe and CRp numerically, we use the operator-split method for dividing the spatial and momentum term.
The fifth-order monotonicity-preserving method is adopted to solve spatial advection term of the equations \eqref{eq:CRe1} and \eqref{eq:CRp1}.
Momentum advection operator can be computed using a second-order piecewise linear construction following \citet{2019MNRAS.488.2235W}.
We use van Leer flux limiter \citep{1977JCoPh..23..276V}.
Velocity divergence, $(\nabla \cdot \bm{v})$, is computed by the center difference method.
Finally, the explicit-solver is used to calculate the momentum diffusion term.
%%% Meshについて
The momentum bins are equally spaced in logarithmic space, as there are 60 bins in  $\gamma_{\rm e} \in [5.0\times 10^1, 5.0 \times 10^6]$ and 90 bins in $\gamma_{\rm p}\beta_{\rm p} \in [5.0 \times 10^{-1}, 5.0\times 10^8]$, for CRe and CRp, respectively.
%The number of bins are 60 and 90 for CRe and CRp, respectively .
%%%
Although CR particles in our simulation do not affect the fluid dynamics, we adopt on-the-fly approach to calculate the diffusion term accurately and anticipate future development.

%  -------------------------- Table -------------------------  %
\begin{table}
  \begin{center}
  \caption{Jets and ICM setup parameters of MHD simulation}
  \label{tab:model}
    \begin{tabular}{c c c}
      \hline\hline
      Jet Kinetic power       & $L_{\rm kin}$     & $2.0\times10^{44}$ [erg ${\rm s^{-1}}$] \\
      Jet thermal power       & $L_{\rm th}$      & $5.4\times10^{43}$ [erg ${\rm s^{-1}}$] \\
      Jet magnetic power      & $L_{\rm mag}$     & $2.0\times10^{42}$ [erg ${\rm s^{-1}}$] \\
      Jet CRp power           & $L_{\rm CRp}$     & $6.0\times10^{42}$ [erg ${\rm s^{-1}}$] \\
      Jet CRe power           & $L_{\rm CRe}$     & $2.0\times10^{42}$ [erg ${\rm s^{-1}}$] \\
      Jet radius              & $r_{\rm jet}$     & 3 [kpc] \\
      Jet Sonic Mach Number   & ${\mathcal M}_{\rm jet}$ & 2 \\ 
      Jet magnetic field      & $B_{\rm \phi,jet}$ & 6.8 $[\mu {\rm G}]$ \\
      Jet plasma beta         & $\beta_{\rm jet}$ & 20  \\ \hline
      ICM temperature         & $T_{\rm ICM}$     & 5 [keV] \\
      ICM number density      & $  n_{\rm ICM}$   & $5\times10^{-3}~{\rm [cm^{-3}]}$ \\ 
      Wind velocity           & $  v_{\rm ICM}$  & $4.0\times10^{2}~{\rm [km~s^{-1}]}$ \\ \hline
      Bending radius          & $  l_{\rm b}$     & 70 [kpc] \\
      Density ratio           & $  n_{\rm jet}/n_{\rm ICM}$ & 0.01 \\
      \hline\hline
  \end{tabular}
\end{center}
\end{table}
%-------------------------------------------------------------------
%----------------------------------------------%
\begin{table}
  \begin{center}
  \caption{Parameters of CR injection}
  \begin{tabular}{cc} \hline \hline
    $N_{\rm e,0}$~[$\rm cm^{-3}$] & $6.0\times10^{-7}$ \\
    $N_{\rm p,0}$~[$\rm cm^{-3}$] & $3.3\times10^{-10}$ \\
    $p$ & 2.1 \\
    $\gamma_{\rm e,min}$ & $5 \times 10^{2}$ \\
    $\gamma_{\rm p,min}$ & $5.0$ \\
    $\gamma_{\rm e,max}$ & $1.0\times 10^{5}$ \\
    $\gamma_{\rm p,max}$ & $1.0\times 10^{5}$ \\ \hline \hline
    \end{tabular}
  \end{center}
  \label{tab:cr_injection}
\end{table}
%--------------------------------------------%

\subsection{Model for turbulence reaccelartion} \label{subsec:reacc.}
%% 素人向け解説の追加 %%
In 3D MHD simulations, large scale vortices cascade down to smaller scales, and then the kinetic energy of the vortex is dissipated numerically when its size is comparable to the MHD grid size.
%is induced by the energy injection from AGN jets,
In actual astrophysical environments, CR particles could be accelerated by interaction with a turbulent eddy whose scale is not fully resolved in numerical simulations.
% Although the efficiency of the stochastic acceleration is strongly related to the physical property of sub-grid scale turbulence, part of turbulence energy should be converted into the energy of CR particles.
From the multi-scale nature of the system, one can presume that the strength of the sub-grid scale turbulence scales with the dissipation energy at each MHD grid. 
Therefore, in this work, we assume that a portion of the turbulence dissipation is converted into the energy of CR particle. 
%%%%%%%%%%%%%%%%%%%%%

% The stochastic acceleration process in turbulence is highly uncertain, and thus
% we adopt the hard sphere approximation, $D_{\gamma \gamma} = K \gamma^2$ for simplicity.
The stochastic acceleration process in MHD turbulence is studied to explain the observed radio emission from galaxy clusters \citep[e.g.,][]{1987A&A...182...21S,2001MNRAS.320..365B}.
It is often assumed that $D_{\gamma \gamma} = K \gamma^2$, so called hard-sphere approximation \citep{2007MNRAS.378..245B,2019ApJ...877...71T}.
Under this assumption, the acceleration timescale $\tau_{\rm acc} \equiv \gamma^2/(4D_{\gamma \gamma})=(4K)^{-1}$  is independent of the CR momenta.
% The acceleration timescale of the hard sphere acceleration, which is regulated by the energy dissipation rate $\dot{u}_{\rm diss}$, can be written as
We assume that the acceleration timescale depends on the energy dissipation rate of turbulence $\dot{u}_{\rm diss}$ in the MHD simulation as
\begin{equation}
    \tau_{\rm acc} =  \frac{u_{\rm e}+u_{\rm p}}{4\eta \dot{u}_{\rm diss}},
\end{equation}
where $u_{\rm e,p}$ are the energy of CRe and CRp, respectively, and $\eta$ is the efficiency of energy conversion from dissipated turbulence to CR particles, which is a parameter in this study.

The energy dissipation rate is frequently discussed in the context of the two-temperature MHD simulation for the hot accretion flow and AGN jets.
This work follows the methodology of those studies \citep{2015MNRAS.454.1848R,2017MNRAS.466..705S,2020MNRAS.493.5761O}, by adopting
\begin{equation}
    \dot{u}_{\rm diss} = \frac{u_{\rm int} - u_{\rm ad}}{\Delta t},
\end{equation}
in each numerical cell.
Here, $u_{\rm int}$, $\Delta t$, and $u_{\rm ad}$ are the internal energy density of the thermal gas, time step of the explicit solver for MHD equations, respectively, and the internal energy density estimated with purely adiabatic evolution, respectively.
For computing the adiabatic evolution, we solve the entropy evolution equation with the MHD equations as
\begin{equation}
    \pdv{t} \qty( \rho s_{\rm gas}) + \nabla \cdot \qty(\rho s_{\rm gas} \bm{v}) = 0,
\end{equation}
where $\rho$ is the density of thermal particles and $s_{\rm gas} = p_{\rm gas} \rho^{-\Gamma}$ is the pseudo-entropy.
We re-calculate $s_{\rm gas}$ at the start of each time steps, and solve the equation by adopting the fifth-order monotonicity-preserving method.
The internal energy density that evolves under the adiabatic process is then computed as
\begin{equation}
    u_{\rm ad} = \frac{s_{\rm gas}\rho^{\Gamma}}{\Gamma-1}.
\end{equation}
As already discussed in section 3.1 of \citet{2017MNRAS.466..705S}, the finite difference and finite-volume approach to solving MHD equations artificially increase the entropy in a grid when a hotter gas and a cooler gas are mixed in the grid.
Therefore, the energy dissipation tends to be overly estimated, especially around contact discontinuity, and the CR energy is also overly estimated in our code.
We include this effect as the uncertainty in the phenomenological parameter $\eta$.

\subsection{Particle injection}
We assume that the jet in our simulation has already experienced several shocks near the launch region.
Those shocks may correspond to radio knots frequently identified in observations.
Thus, we inject CRe and CRp into the jet at the jet injection area of the simulation.
Since the sonic Mach number of our jets is 2, those jets do not induce strong shock waves.
% which is much higher than a critical sonic Mach number for stochastic shock drift acceleration $\mathcal{M}_{\rm s,crit} \approx 2.25$ \citep{2018ApJ...864..105H}.
Therefore, we neglect additional injection of CR in our simulations. 
% That is consistent with observations, that there are no bright radio knots, namely strong shocks, in tailed-region.

The energy distribution is assumed to be a single power-law with an exponential cut-off as
%at launching region:
\begin{equation}
    Q_{\rm s} = N_{\rm s,0} \gamma^{-p} \exp(-\gamma/\gamma_{\rm s,max}) ~~~(\gamma > \gamma_{\rm s,min}),
\end{equation}
where the subscript $s$ refers to electrons and protons ($s = {\rm e,p}$) with $N_{\rm s,0}$, $p$, $\gamma_{\rm s,min}$, and $\gamma_{\rm s,max}$ are the model parameters, whose values are listed in Table \ref{tab:cr_injection}.
Under this parameter set, the CR proton-to-electron number ratio is 3.5 at 10 GeV.

As mentioned above, the acceleration timescale depends on the total CR energy in our sub-grid model.
Injection parameters of CRe and CRp, therefore, influence our results.
In this work, we simply use the equipartition condition, $u_{\rm e} \approx u_{\rm p} \approx u_{\rm mag}$, at the jet injection point.
% The choice of $p=2.1$ is consistent with radio observations, which indicate that the radio spectral index ($\nu^{\alpha}$) of the region where close the AGN core for head-tail galaxies are $\alpha \approx -0.5$.
The choice of $p=2.1$ ensures that the radio spectral index ($\nu^{\alpha}$) in the region near the AGN core for head-tail galaxies is $\alpha \approx -0.5$, which is consistent with radio observations \citep{1976ApJ...203..313P}.
From the radio and X-ray observations of radio jets, the energy density of the magnetic field and CRe can be estimated, and these energies are comparable, though the energy density of CRe can be slightly larger than that of the magnetic field \citep{2002ApJ...581..948H, 2004ApJ...612..729H}.
Since low-energy CRe ($\gamma_{\rm e} \lesssim 500$) rapidly loses energy by Coulomb interactions \citep{1999ApJ...520..529S}, we use $\gamma_{\rm e,min} = 500$.
However, it is difficult to constrain the ratio of energy in CRp to CRe from observations, while a significant contribution of non-radiating particles (CRp and/or thermal particles) to inflate the radio lobe of the FR-I jets is needed \citep{2000MNRAS.319..562H,2014MNRAS.438.3310C}.
Note that a larger $\eta$ value is required to obtain the same acceleration efficiency of CRe if $u_{\rm p} \gg u_{\rm e}$.

\subsection{Non-thermal emission}
% In this work, we model the radio synchrotron emission and the X-ray inverse Compton emission by integrating each emissivity along lines of sight the entire domain.

We calculate the intensities $I(\nu)$ of electromagnetic waves and neutrinos by integrating emissivities along lines of sight throughout the entire simulation domain.
In this work, we consider lepton emission with synchrotron and inverse Compton scattering, and hadronic emission due to interactions of CRp with thermal protons.

The synchrotron emissivity from CRe (in optical thin limit) is given by
\begin{equation}
    \mathcal{J_{\nu}}^{sync} = \frac{1}{4 \pi} \int P_{\nu}(B_{\perp}, \gamma) N_{\rm e}(\gamma) d\gamma,    
\end{equation}
where $P_{\nu} (B_{\perp}, \gamma)$ and $B_{\perp}$ are the specific emissivity of a single electron by synchrotron radiation and the strength of the magnetic field perpendicular to the lines of sight, respectively.
To reduce the computational effort, we use the fitting formula of \citet{2013RAA....13..680F} for the specific emissivity.
After calculating the surface brightness maps, it is smoothed by the Gaussian convolutions parameterized by a beam size.

A radio spectral index is computed using two radio maps at different frequencies $\nu_1$ and $\nu_2$ as follows:
\begin{equation}
    \alpha_{\nu_1-\nu_2} = -\frac{\log_{10} \{ I(\nu_2)/I(\nu_1) \}}{\log_{10} (\nu_2/\nu_1)}.
\end{equation}

For the emissivity of inverse Compton radiation with CMB photons, we use the formula given in \citet{1996ApJ...463..555I}:
\begin{equation}
    \mathcal{J_{\nu}^{IC}} = \frac{h}{4\pi} \varepsilon q(\varepsilon),~~~ \nu = \frac{m_{\rm e}c^2}{h} \varepsilon,
\end{equation}
\begin{equation}
    q(\varepsilon) = \int d\varepsilon_0 n_{\rm ph}(\varepsilon_0) \int d\gamma N_{\rm e} (\gamma) C(\varepsilon, \gamma, \varepsilon_0),
\end{equation}
where $\varepsilon m_{\rm e}c^2$, $\varepsilon_{0} m_{\rm e}c^2$, $h$ and $n_{\rm ph}(\varepsilon_0)$ are the energy of CRe, the target photon energy, the Planck constant and the number density of CMB photons per energy interval, respectively.
The function $C$ is called the Compton kernel \citep[see equation (44) in][]{1968PhRv..167.1159J}.
We assume that $n_{\rm ph}(\varepsilon_0)$ is a blackbody spectrum with the temperature $T_{\rm CMB}$.

For computing the emissivities of hadronic gamma-ray and neutrinos from the $pp$ collision, we use the numerical code of \citet{2021ApJ...922..190N}.  
This code uses the approximate expression for the spectra of pions and neutrinos given in \citet{2006PhRvD..74c4018K} and the inclusive cross section for neutral and charged pion productions from \citet{2006ApJ...647..692K,2007ApJ...662..779K}.

\section{Results} \label{sec:Results}

\subsection{Overview of Simulations}
\begin{figure}[h]
    \centering
    \includegraphics[width=0.5\textwidth]{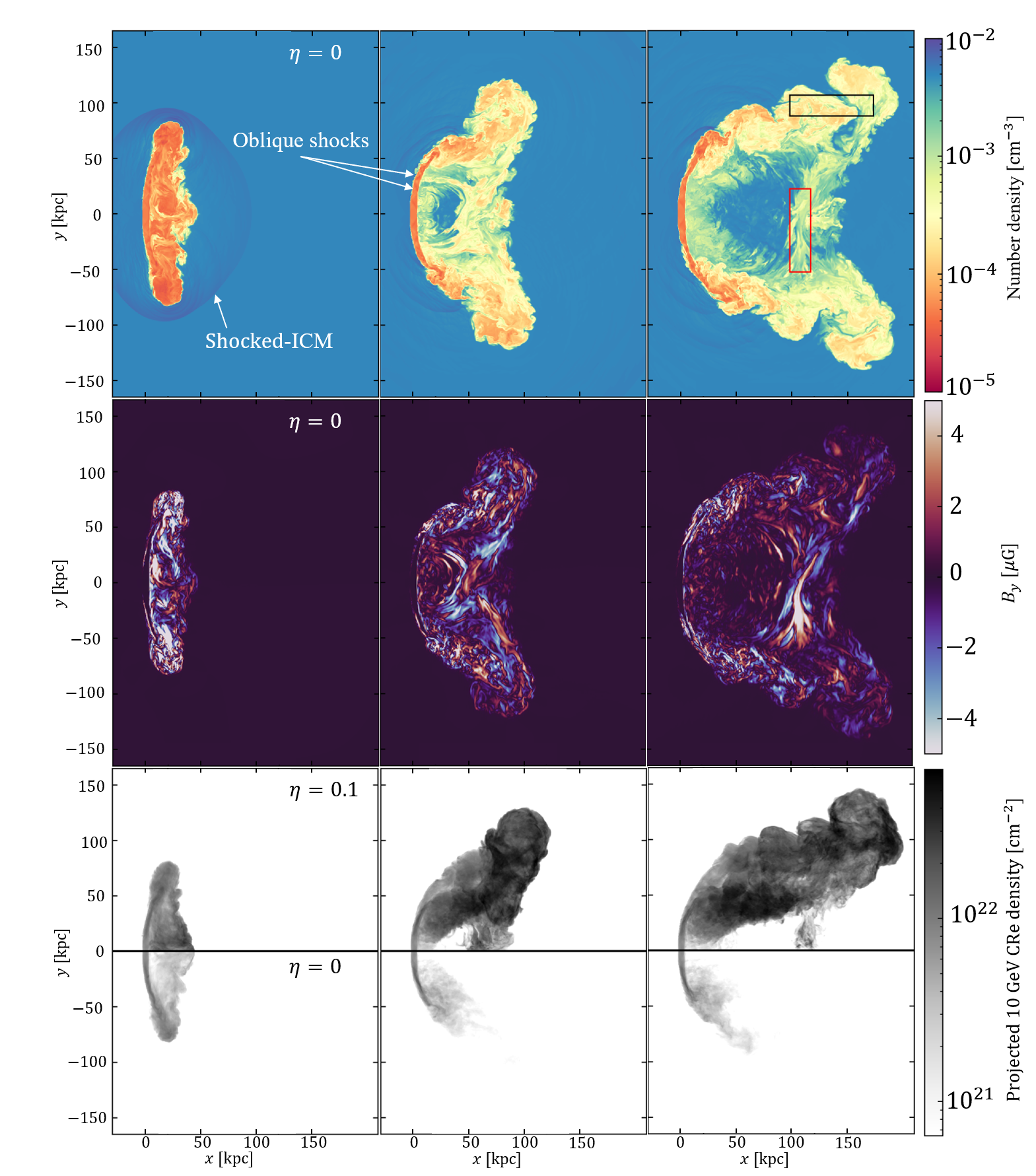}
    \caption{Number density (top) and the magnetic field strength (middle) maps ($z=0$ kpc) at $t=96.6,$ 193.2, and 316.3 Myr (from left to right).  
    The rectangles indicate the regions to compute the characteristic scales of magnetic field for the tail (black) and the threads (red) (see section \ref{sec:diffusion}).
    The bottom panels show maps of 10 GeV CRe column density, $\int N_{\rm e,10GeV} dz$, for $\eta = 0$ (lower half) and 0.1 (upper half).} 
    \label{fig:ro_by}
\end{figure}

\begin{figure*}[h]
    \centering
    \includegraphics[width=0.8\textwidth]{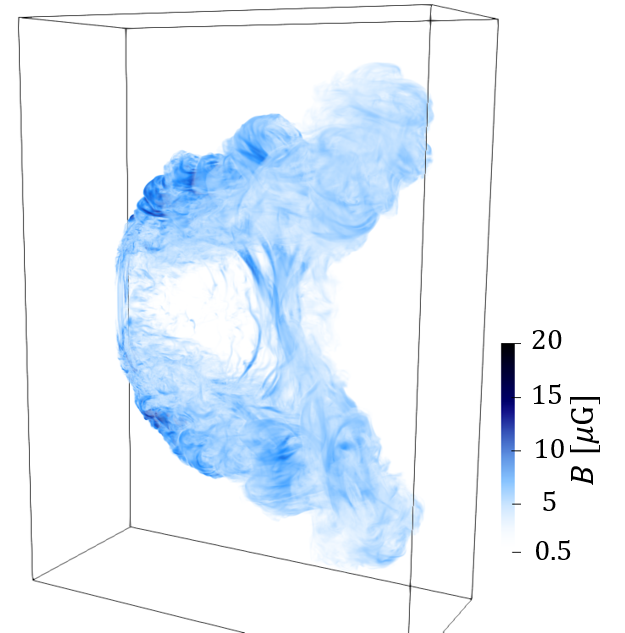}
    \caption{3D render of magnetic field strength at $t = $ 316.3 Myr.  
    The box size is $240 \times 328 \times 104$ ${\rm kpc^3}$.
    % The elongated flux tubes, which link the both tails, can be seen.
    }
    \label{fig:3D_thread}
\end{figure*}

% First, we report the dynamics of the head-tail galaxy.
The overall morphology and dynamics of the jets are similar to that discussed in previous simulations \citep{1992ApJ...393..631B,2009AIPC.1201..259P,2019ApJ...884...12O}.
In Figure \ref{fig:ro_by}, we show snapshots of the number density, the magnetic field strength, and the projected 10 GeV CRe density at three different times $t=96.6$, 193.2, and 316.3 Myr.
At early times, one can see the weak forward shock propagating into the ICM.
The jet flows, which are heated up through the reverse shock (termination shock), decelerate and expand transversely.
If there is no wind in the ambient ICM, these jets flow back symmetric along the jet axis and form 'cocoon' \citep[e.g.,][]{1982A&A...113..285N}.
Since the backflow material is hot and light, the fluid is advected toward in the downwind direction.
As a result, the jet material forms a 'U' shape.
Here, the important feature is that the backflowing plasma is connected to the opposite side of the jet tips.
This phenomena is associated with the formation of magnetic threads in the later phase.

%-------------%
% After the jet flow change direction at the bending radius, $l_{\rm b} \sim 70$ kpc, by $\sim 90^{\circ}$ (see the middle panel of Figure \ref{fig:ro_by}), we can see the episodic jet flapping and disruption events via Kelvin-Helmholtz (KH) instability, which are also seen in \citet{2019ApJ...884...12O}.
% The weak oblique (recollimation) shocks are formed along the jets, and the jet velocity decreases through these shocks.
% Beyond the point of maximum curvature along the jet axis, the flow  decrease to sub-sonic velocities, and the turbulence are driven due to the growth of the KH instability that occur between the lobe and the ICM by their velocity shear.
% Hereafter, we call this diffusive region as the 'tail', and the subsonic turbulence are developed in the tail.
%--------------%
After the jet flow changes direction at the bending radius, $l_{\rm b} \sim 70$ kpc, by $\sim 90^{\circ}$ (see the middle panel of Figure \ref{fig:ro_by}), sub-sonic and super-Alfvenic turbulence are developed due to the growth of the Kelvin-Helmholtz (KH) instability between the lobe and the ICM.
The weak oblique (recollimation) shocks are formed along the jet beams, and the jet velocity decreases through these shocks.
The sonic Mach number of these shocks is about 2 so that the reveres shock may not lead sufficient shock acceleration.
Beyond the point of maximum curvature along the jet axis, the flows are decelerated to sub-sonic velocities.

% The pressure in the tail is greater than that in the surrounding ICM, thus causing the tail expansion while the ICM is mixed.
Since the pressure of the tail is greater than that of the surrounding ICM, the tail expands being mixed with ICM.
One can see large vortices, whose diameter is about 30 - 50 kpc, in both the tails at $t = $ 316.3 Myr (the right panel of Figure \ref{fig:ro_by}).
% This interval of the vortex formation is roughly same as the interval of jet flapping events.
% Thus, the number of the large vortex is written as $N_{\rm vor} \sim t_{\rm age}/\tau_{\rm eddy}$, where $\tau_{\rm eddy}$  is a characteristic eddy turnover time of largest scale.
According to the discussion in \citet{2019ApJ...884...12O}, the eddy turnover timescale can be roughly estimated as $\tau_{\rm eddy} \sim l_{\rm t}/v_{k, {\rm min}} \sim l_{\rm b}/v_{\rm w} \sim$ 80 Myr.
% Note that we confirm that the velocity in the tails is close, but slight higher than, the wind velocity.

% Magnetic field について
In our simulations, the magnetic field is provided only from the jet launching region.
Since the plasma-$\beta$ for the jet, $\beta_{\rm jet} = 20$, is high, the jet morphology is not affected by the magnetic force.
% But, the information of magnetic field is directly related to synchrotron radiation and cooling for CRe.
We find that the magnetic field gradually decreases along the tail because the tail is expanding, and the diffusion and dissipation of the magnetic field occur at the boundary between tails and ICM.
Note that in simulations the diffusion and dissipation of the magnetic field are numerically induced.
% We assume the un-magnetized ICM, and this assumption may tend to underestimate the magnetic field strength in the tails.
We confirm that although the symmetrical boundary condition at $y = 0$ does not affect the dynamics in the simulated results significantly by comparing the results for the symmetric and non-symmetric cases.

The bottom panels in Figure \ref{fig:ro_by} show the effect of reacceleration for CRe.
For $\eta = 0.1$, the radio-emitting CRe, whose typical energy is 10 GeV, appear even at the end of the tail.
The CRe are accumulated around the region of $|y| < 50$ kpc, where the small-scale KH vortices are well developed.
However, the magnetic field of this region is lower than that of the jets (see Figure \ref{fig:3D_thread}).
On the other hand, for $\eta = 0$, there is no 10 GeV CRe in the tail. 
Since the cooling time for 10 GeV at $B = 8$ $\mu$G is $\sim$ 170 Myr, the first injected CRe are already cooled at $t = 193.2$ Myr (see the lower middle column in Figure \ref{fig:ro_by}).

\subsection{Formation of magnetic threads}
Figure \ref{fig:3D_thread} shows magnetic field threads connecting the two tails.
Similar structures are also formed in the simulations of \citet{2019ApJ...884...12O} and \citet{2022arXiv220604757N}, but the threads in our simulations are thinner and collimated, because of the high-spatial resolution scheme.
The two large threads have opposite $y$-direction magnetic field.
The radius and length of these threads are about 10 kpc and 150 kpc at $t = 316.3$ Myr, respectively.
The threads have relatively strong magnetic field, compared to that of the tail region.
% At the initial stage of jet dynamics, the backflowing plasma is to reach and connect the other side of jet tips.
As we have pointed out, the backflowing materials at the initial stage are the origin of the magnetic threads.
The threads dynamically evolve with the wind, i.e., the backflowing materials are advected by the wind forming the threads.
% the threads are advected by the wind.

\begin{figure}[h]
    \centering
    \includegraphics[width=0.5\textwidth]{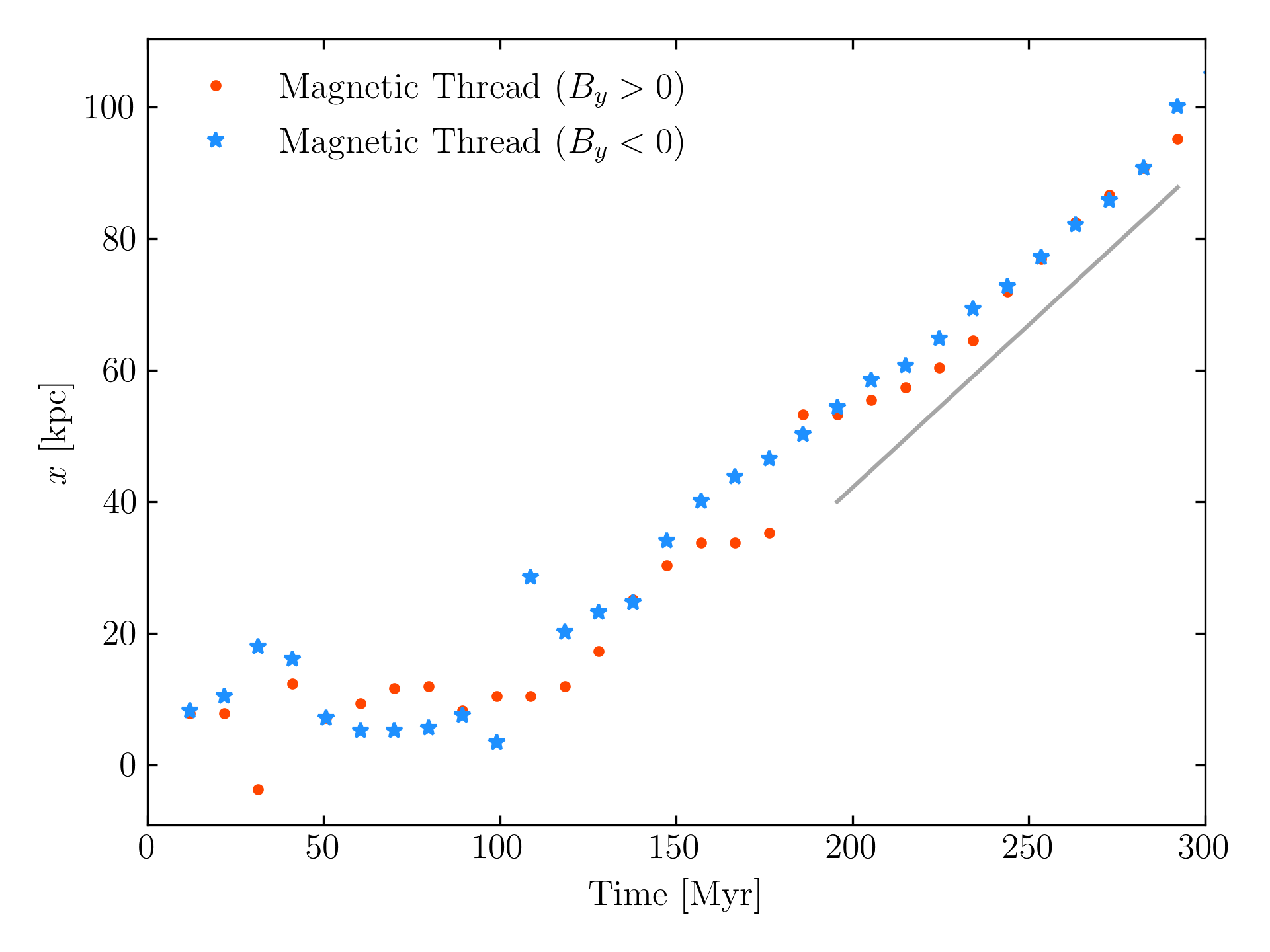}
    \caption{The time evolutions of the positions of the two magnetic threads at $y = 0$.
    The gray line shows the wind motion, $x = v_{\rm ICM} t$
    }
    \label{fig:filament}
\end{figure}

To clarify this point, we examine the time evolutions of the positions of the two threads. 
% Calculating the position for magnetic threads, firstly 
The average $y$-direction magnetic field in each grid point $(i,j,k)$ is calculated as
\begin{equation}
    \bar{B}_{y} = \frac{1}{11^2} \sum^{i+5}_{i' = i-5} \sum^{k+5}_{k' = k-5} B_{y,(i', j, k')},
\end{equation}
by averaging magnetic field $\pm 10$ grids in $x$ and $z$ directions.
Then, we define the positions of the rich magnetic threads as the position where $|\bar{B}_{y}|$ is the maximum at $y=0$ plane.
As shown in Figure \ref{fig:filament}, the advection velocities of both the threads are almost the same as the wind velocity after $t = 100$ Myr.

\subsection{CRe and CRp energy evolution} \label{subsec:CR energy}

\begin{figure*}[t]
\begin{tabular}{cc}
\begin{minipage}{0.48\textwidth}
\centering
\includegraphics[width=1.0\textwidth]{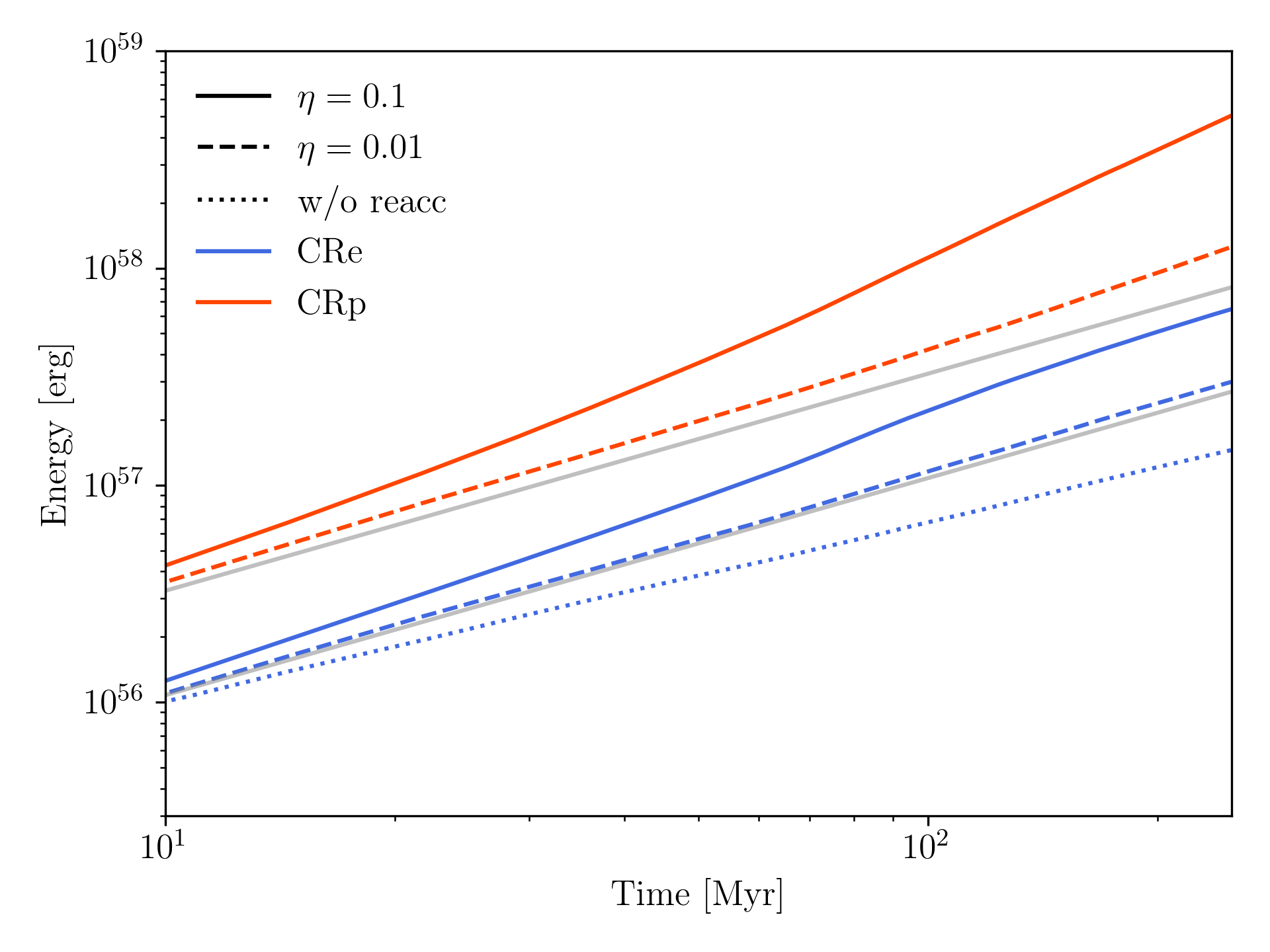}
\end{minipage} &

\begin{minipage}{0.5\textwidth}
\centering
\includegraphics[width=1.0\textwidth]{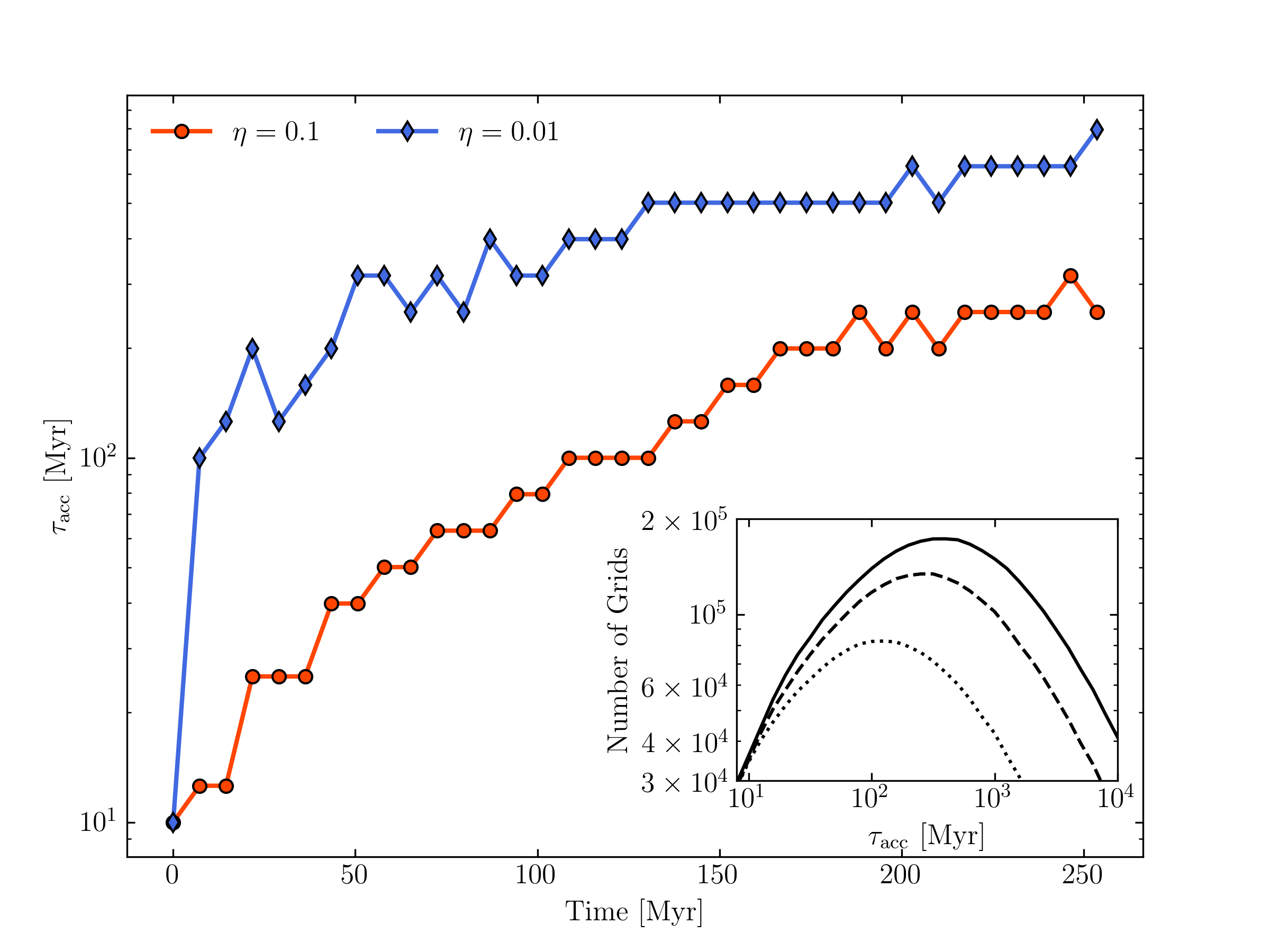}
\end{minipage}

\end{tabular}
\caption{Left: Time evolutions of the energies of CRe (blue) and CRp (red) in the systems for $\eta = 0$ (dotted), $0.01$ (dashed), and $0.1$ (solid). The gray solid lines are the expected growth of the CRe and CRp energies without energy gain and loss processes. Right: Time evolutions of the mode value of the acceleration timescales for $\eta = 0.01$ (blue diamond)  and $0.1$ (red circle). The inlet shows the distributions of the acceleration timescales at $t =$ 96.6 (dotted), 193.2 (dashed), and 316.3 (solid) Myr for $\eta = 0.01$.
}
\label{fig:CR_evolution}
\end{figure*}

% \begin{figure}[h]
%     \centering
%     \includegraphics[width=0.5\textwidth]{fig/tacc_tevo.png}
%     \caption{Time evolution of mode values of acceleration timescale for $\eta = 0.01$ (blue diamond)  and $0.1$ (red circle). The inlet shows the distribution of acceleration timescale at $t =$ (dotted), (dashed), and (solid) for $\eta = 0.01$.}
%     \label{fig:tacc_tevo}
% \end{figure}

\begin{figure}[h]
    \centering
    \includegraphics[width=0.5\textwidth]{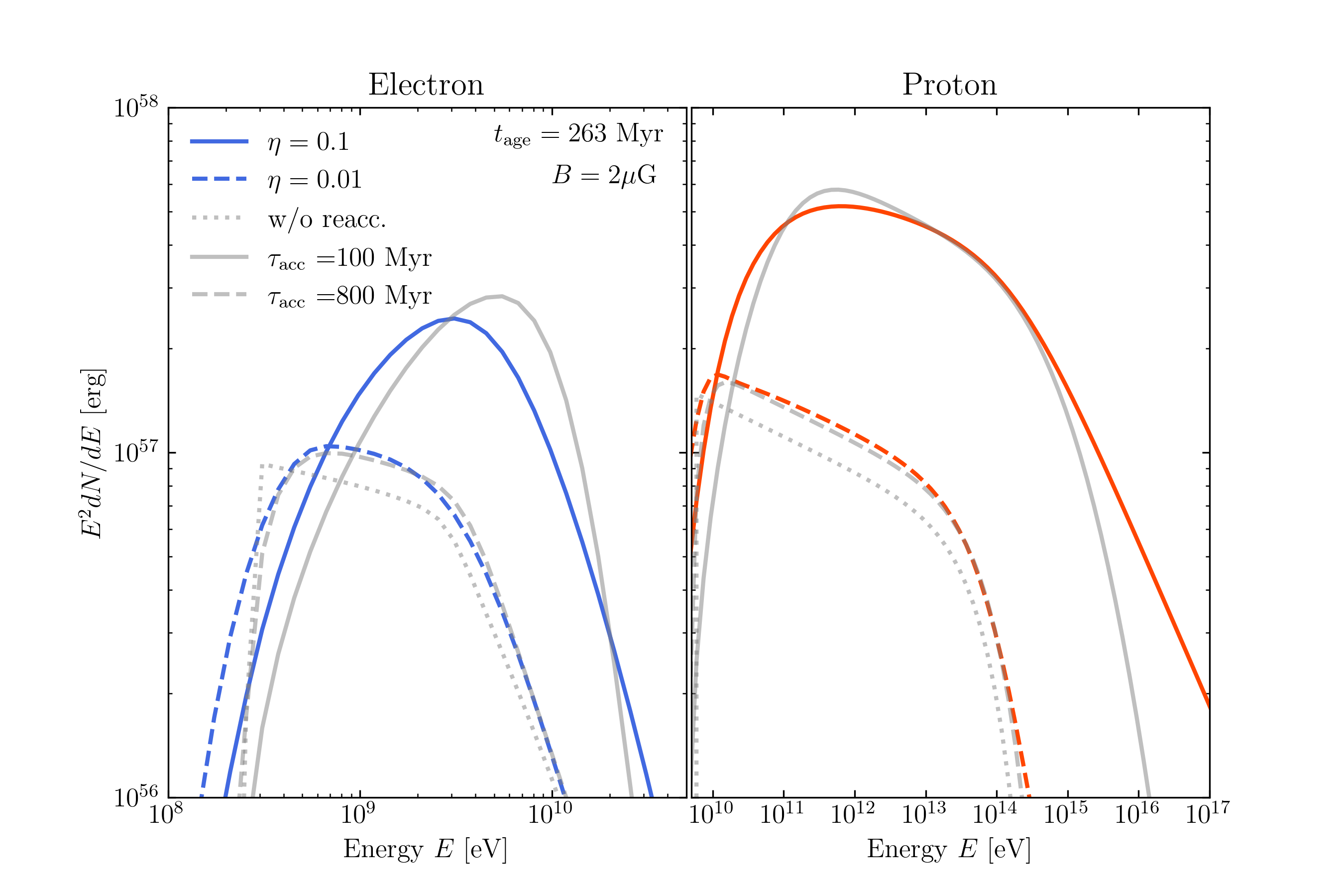}
    \caption{The CRe (left) and CRp (right) energy distributions integrated over the whole region for $\eta = 0.01$ (dashed) and 0.1 (solid) at $t = 316.3$ Myr. The gray lines show the spectra for the one-zone model with $\tau_{\rm acc} =$ 100 (solid), 800 (dashed), and $\infty$ (dotted) Myr, respectively.}
    \label{fig:CR_spectrum}
\end{figure}

The CRp and CRe energies evolve as shown in Figure \ref{fig:CR_evolution}.
As mentioned in section 2.3, the effective value of $\eta$ is larger than the values we set.
% The CR energies gained by adiabatic compression and 2nd-order Fermi acceleration since our simulations ignore particle injections into the diffusive shock acceleration (DSA) like acceleration process. 
Our simulations show that 3 \% and 30 \% of the jet kinetic energy, $E_{\rm jet} = L_{\rm kin} t \sim 2.0 \times 10^{59}$ erg, is converted into the CR particles energy at 316.3 Myr for $\eta = 0.01$ and $0.1$, respectively.
We check that the effective efficiencies $\eta^{\rm eff} \equiv \Delta E_{\rm CR}/E_{\rm diss}$, where $\Delta E_{\rm CR}$ and $E_{\rm diss}$ are respectively the energy gain of CR particles and the total dissipation energy, are $6.5 \times 10^{-3}$ and $6.6 \times 10^{-2}$ for $\eta = 0.01$ and 0.1 at 316.3 Myr, respectively. 
This implies that the about 35 \% of the gain energy is lost via radiation and adiabatic processes.

The acceleration timescale for CR particles is not constant, and depends on the total CR energy density and dissipated energy at each position.
In Figure \ref{fig:CR_evolution}b, the acceleration timescales approximately follows a log-normal distribution.
The mode value of the acceleration timescales monotonically increases and saturates at several hundred Myrs for both the models.
There are two reasons for this trend.
First, the volume of the tail increases with time, and also the turbulence decays along with the tail.  
As a result, the region with a lower dissipation rate expands.
Since the flows around the bent region are supersonic, the dissipated energies are much larger than that around the tail region. 
% (2) the larger the energy density of CR particles is, the longer the acceleration timescale becomes.
Second, as the energy density of CR particles inceases, the acceleration timescale becomes longer.

% CRe
Since CRe lose its energy by radiations, the growth rate of the CRe energy is lower than that of the CRp energy.
For the case of $\eta = 0.01$, the radiative energy loss is comparable to the energy gain due to reacceleration (see Figure \ref{fig:CR_evolution}), as the cooling timescale for  GeV CRe is about 100 Myr in $\sim  \mu$G magnetic field.
Meanwhile, the energies of CRe and CRp for $\eta = 0.1$ drastically increase because the jet age and the acceleration time are always comparable.
The CRp energy of $\eta = 0.1$ is about 10 times larger than that without reacceleration at $t =$ 316 Myr, while the CRe energy is slightly larger than that without reacceleration, because of the radiative cooling.

The CRe and CRp energy distributions integrated over the whole region are shown in Figure \ref{fig:CR_spectrum}.
For comparison, we calculated the time evolution of CR particles with a one-zone model, where we neglect the advection and the adiabatic terms in equations \eqref{eq:CRe1} and \eqref{eq:CRp1}.
% which are given as follows:
% \footnotesize
% \begin{eqnarray}
%     \label{eq:CRe}
%     \pdv{N_e}{t}  &=&  
%      \pdv{\gamma}\qty[ - N_e \dot{\gamma}_{\rm cool}] \nonumber \\ 
%     &-& \pdv{\gamma} \qty[\frac{N_e}{\gamma^2} \pdv{\gamma}\qty(\gamma^2 D_{\gamma \gamma})]+\pdv[2]{\gamma}\qty[D_{\gamma\gamma}N_e] + Q_{\rm e,inj}
% \end{eqnarray}
% \begin{eqnarray}
%     \label{eq:CRp}
%     \pdv{N_p}{t}  &=& - \pdv{\gamma} \qty[\frac{N_p}{\gamma^2} \pdv{\gamma}\qty(\gamma^2 D_{\gamma \gamma})]+\pdv[2]{\gamma}\qty[D_{\gamma\gamma}N_p] + Q_{\rm p,inj}.
%     \label{eq:one-zone}
% \end{eqnarray}
% \normalsize
We adopt a constant magnetic field 3 $\mu$G and thermal electron density of 0.01 ${\rm cm^3}$, and the acceleration times of 100 and 800 Myr for $\eta = 0.01$ and 0.1. 
The results are shown by the gray lines in Figure \ref{fig:CR_spectrum}.
For $\eta = 0.01$, the simulation and the one-zone model have similar energy distributions of CRe and CRp.
While most of the acceleration timescales are longer than the age of the jet during the simulation, a small fraction of CRe accelerate very efficiently.
In contrary to this, the energy distributions of both CRe and CRp for $\eta = 0.1$ are significantly different from that of the one-zone model.
At higher energies, the spectra in the simulations is much harder than those for the one-zone model (see Figure \ref{fig:CR_evolution}).
Differently from the one-zone model, the scattered values of the energy dissipation rate in the MHD simulations broaden the CR spectra. In low-dissipation regions, the inefficient acceleration leads to the lower peak of the CR spectra. Alternatively, high-dissipation regions are the main sites where CRs in the high-energy tail are accelerated.

% \begin{Figure}[h]
%     \centering
%     \includegraphics[width=0.5\textwidth]{fig/jet_dissipation_tevo_linear.png}
%     \caption{ Time evolution of the energy converted into CR particles for $\eta = 0.01$ (dotted) and $0.1$ (dashed). The evolution of total jet energy is plotted as black solid line.
%     For $\eta = 0.01$ and $0.1$, 3 \% and 30 \% of total jet energy is converted into CR particles energy at 316.3 Myr, respectively.}
%     \label{fig:jet_diss}
% \end{figure}

\section{Radiation from head-tail galaxy} \label{sec:emission}
In this section, we discuss leptonic and hadronic emissions from our models.
For calculating fluxes, the source luminosity distance is assumed as $D = $ 70 Mpc, which is roughly the same as the distance of NGC 1265 in the Perseus Cluster.
The emissions are calculated by using the snap shot data at $t = 316.3$ Myr.
We assume that the tails lie in the plane of the sky for simplicity.

\subsection{Radio emission}
\begin{figure}[h]
    \centering
    \includegraphics[width=0.5\textwidth]{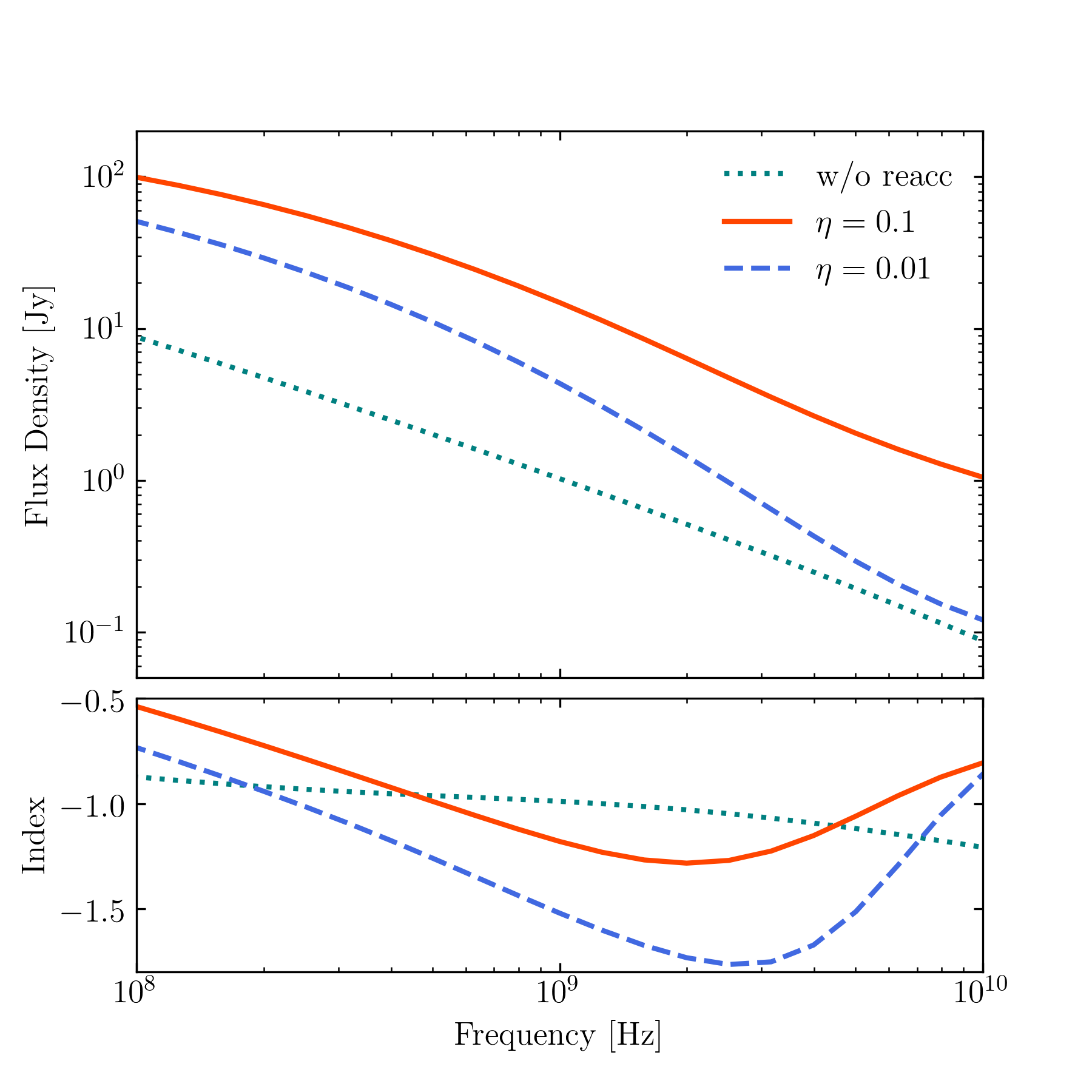}
    \caption{
    Integrated radio spectra for $\eta = 0$ (green dotted), 0.01 (blue dashed), and 0.1 (red solid).
    }
    \label{fig:radio_integrated}
\end{figure}

The integrated synchrotron flux densities are shown in Figure \ref{fig:radio_integrated}.
In the model without reacceleration, the radio spectrum is almost a single power-law with an index of $\alpha \sim -0.9$.
The index is slightly harder than the simple expectation from the continuous injection model, whose electron spectrum, $N(\gamma) \propto \gamma^{-p+1}$, i.e., $\alpha \sim -1.05$.
That difference may come from the effect of adiabatic compression and the Coulomb loss.
In contrast to this, the spectral indices for the models with reacceleration become softer as frequencies get higher.
% The radio spectrum for the model with reacceleration have steeper spectrum around GHz frequencies than that without reacceleration.
Above 5 GHz the radio spectra become harder reflecting the curved electron spectrum. 
% \citet{2017AJ....154..169S} reported integrated flux densities for seven head tail radio sources.
% Although we can see the softening of the spectrum, there is a shortage of radio data to discuss more precisely.

% The distributions of 150 MHz radio luminosity and size for head-tail galaxies, are reported in \citet{2019MNRAS.488.2701M}.
The physical sizes of our models are $\sim$ 200 kpc at the time in our mock observation, and the 150 MHz luminosities are $L_{\rm 150} = 2.1\times 10^{32},~6.0 \times 10^{32}$, and $1.5 \times 10^{33}$ erg/Hz for $\eta =$ 0, 0.01, and 0.1, respectively.
From the LOFAR Two-Metre Sky Survey, $L_{\rm 150}$ is in the ranges between $10^{31}$ and $10^{33}$ for the observed head-tail galaxies, whose physical size are about 200 kpc \citep{2019MNRAS.488.2701M}.
Thus, the model for $\eta = 0.1$ corresponds to the most luminous radio source.
% Those values roughly agree with the observed typical values \citet{2019MNRAS.488.2701M}.

\begin{figure*}[h]
    \centering
    \includegraphics[width=1.0\textwidth]{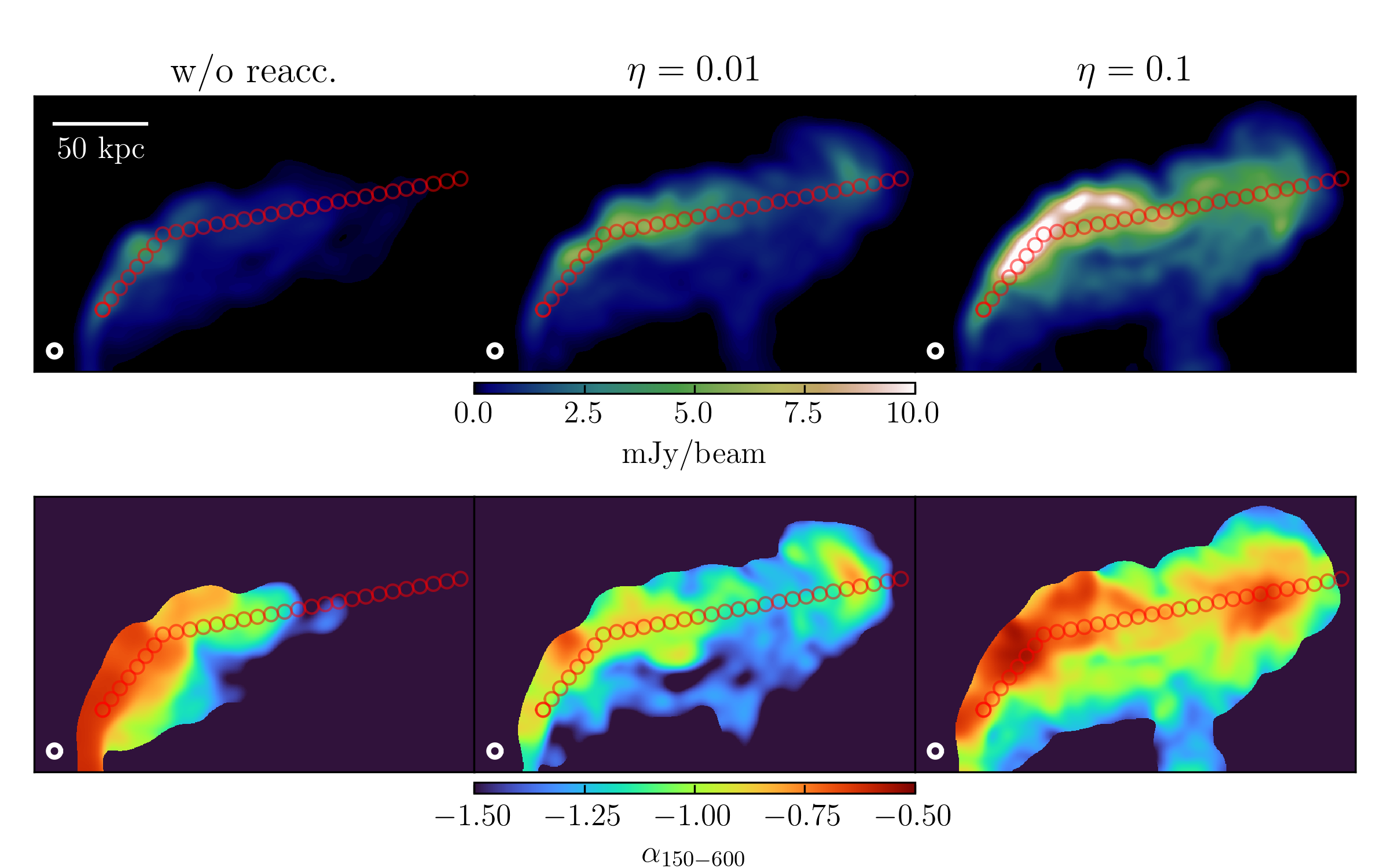}
    \caption{
    \textbf{top}: 300 MHz radio maps for $\eta =$ 0, 0.01, and 0.1 (from left to right). The source luminosity distance is assumed as $D =$ 70 Mpc. The beam size, 7'' $\times$ 7'', is shown in the bottom left corner. Red circles on the radio jet indicate the position where the fluxes are estimated in Figure \ref{fig:radio_flux_distance}.
    \textbf{bottom}: Spectral index maps derived from the radio data obtained at 150 - 600 MHz.
    }
    \label{fig:radiomap}
\end{figure*}

\begin{figure*}
  \begin{minipage}{0.5\hsize}
    \includegraphics[width=1.0\textwidth]{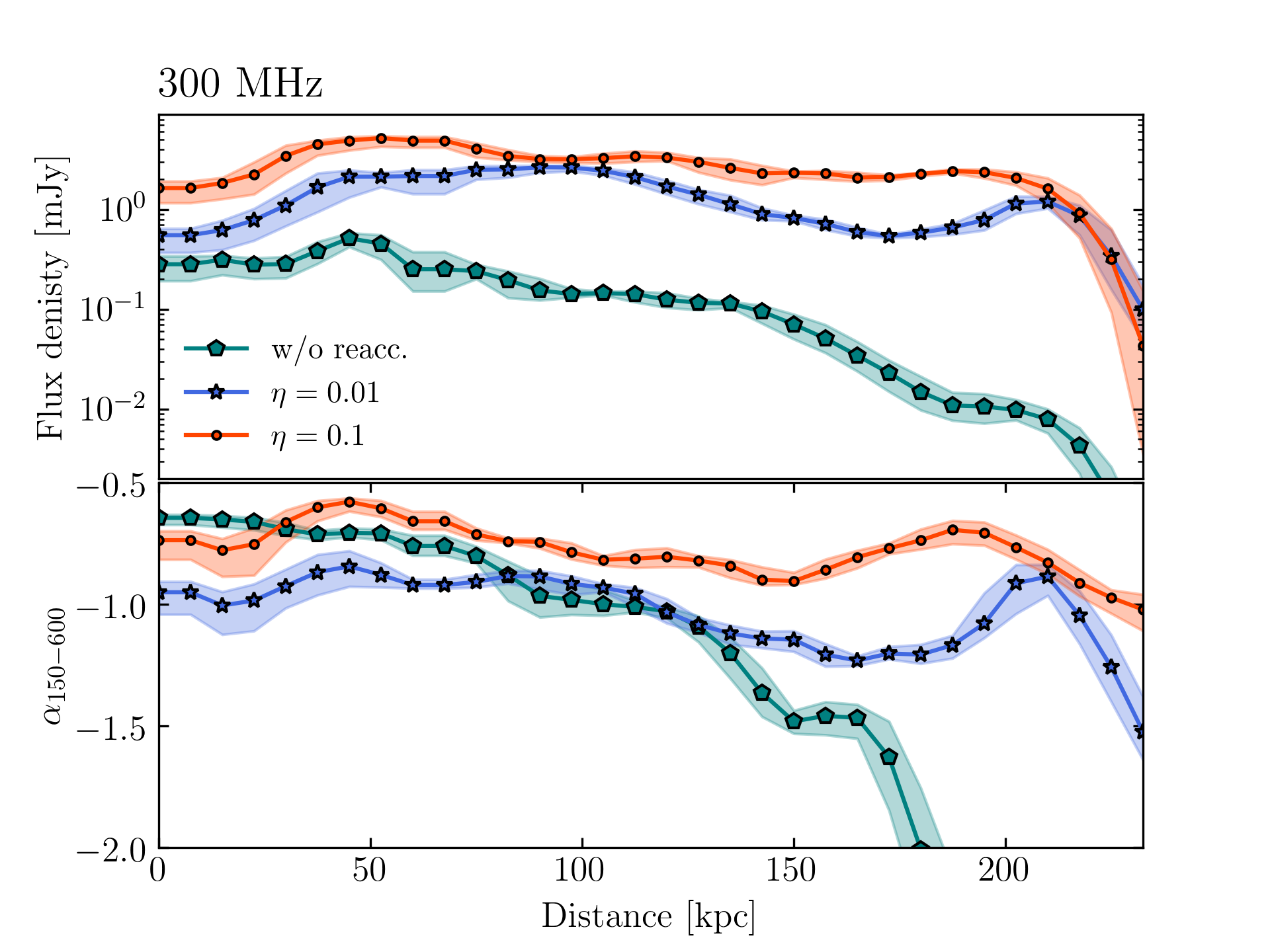}
  \end{minipage}
  \begin{minipage}{0.5\hsize}
    \includegraphics[width=1.0\textwidth]{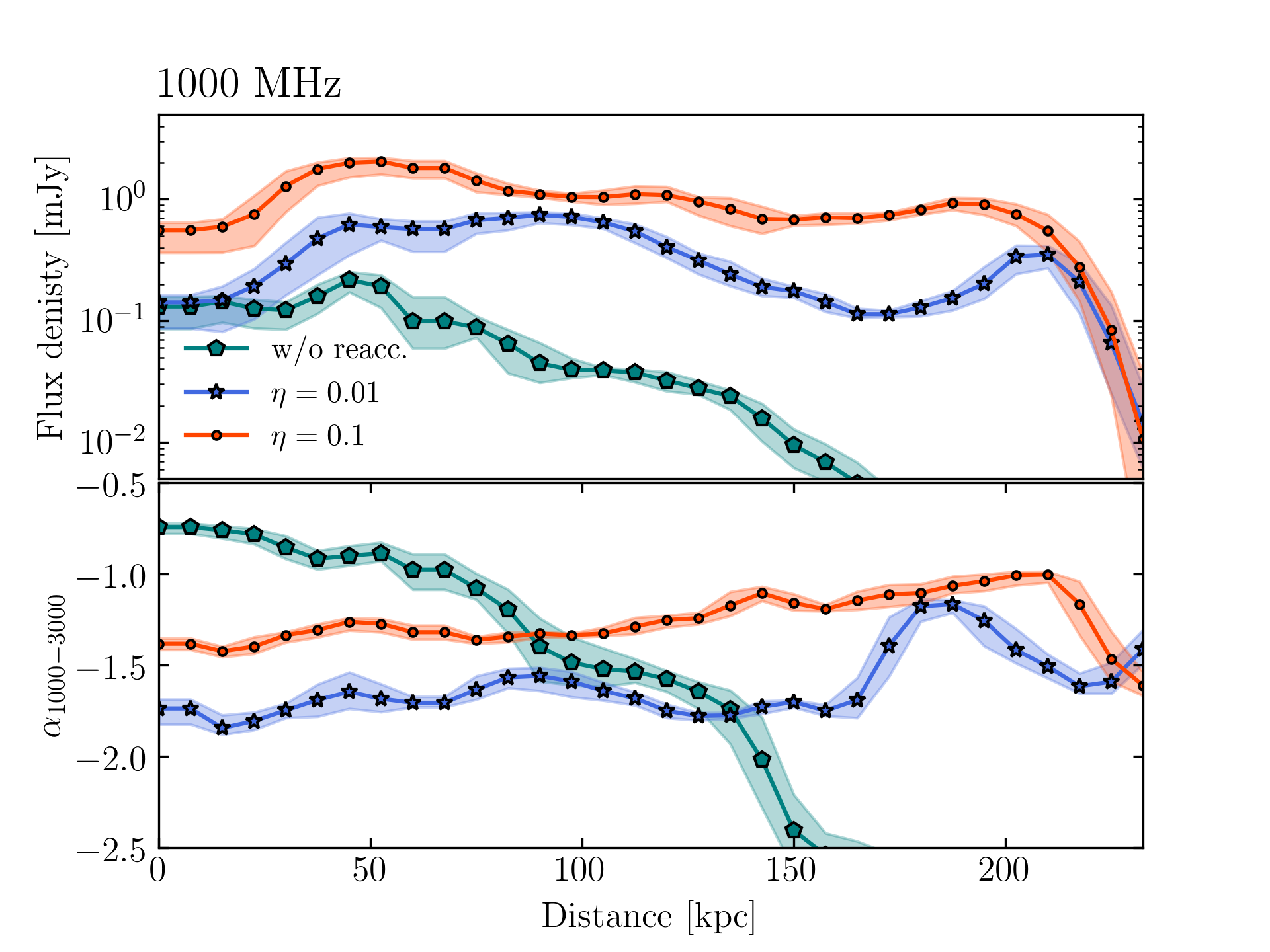}
  \end{minipage}
  \caption{ 
  Flux densities (top panels) and spectral indices (bottom panels) profiles along the radio jets indicated in Figure \ref{fig:radiomap} for $\eta = 0$ (green hexagons), 0.01 (blue stars), and 0.1 (red circles). 
  }
  \label{fig:radio_flux_distance}
\end{figure*}

The 300 MHz radio maps and the spectral index map derived from the radio data at 150 - 600 MHz in our simulations are shown in Figure \ref{fig:radiomap}.
We also show their profiles across the jets in Figure \ref{fig:radio_flux_distance}.
The radio emission is most prominent at the end of the bending, where the magnetic field strength is high (see also Figure \ref{fig:ro_by} and \ref{fig:3D_thread}).
Because of radiative cooling and adiabatic expansion, the brightness in the model without reacceleration is dark in the tail region.
Although a large amount of CRe is accumulated in the tail region, such low-energy electrons do not radiate in the radio band.
% The radio intensity and the spectral index monotonically decreases toward to the tail in the case of no reacceleration.
% The magnetic field and the density of CRe are weak around the tail region due to expansion and gas mixing.
% So that it is difficult to observe radio emission from the model without reacceleration in spite of the low frequency regime. 
This behavior is the same as ones seen in the pure-aging scenario, that is inconsistent with radio observations \citep{1973A&A....26..423J}. 

In the presence of the reacceleration, the radio intensity in the tail region is so high that recent instruments can detect.
The radio flux and spectral index do not change drastically.
Those behaviors are consistent with some head tail galaxies \citep{1976ApJ...203..313P,1975A&A....38..381M,1986ApJ...301..841O,1998A&A...331..475F,2021MNRAS.508.5326M}.
The profiles of the radio flux and the spectral index for $\eta = 0.01$ and 0.1 have similar trends, while the efficiency of the reacceleration affects the normalization.
Thus, it may be difficult to determine the value of $\eta$ from the radio intensity distribution.
Meanwhile, the spectral index offers a hint.
Recent radio observations show that the spectral index for hundreds of MHz frequency range is -1.0 or less in the tails \citep{2017AJ....154..169S,2020MNRAS.499.5791G}.  
Therefore, the spectral index fro $\eta = 0.1$ may be too hard. 
% Interestingly, we can see the bump of radio flux and the flattening of the spectral index at the tip of jets.
% This is because the accumulated CRe at this region can be accelerated for a long time.
% While the values of radio flux and spectral index are different for the reacceleration efficiency, it would be difficult to estimate it from the radio data.

% \citet{2020A&A...636L...1R} showed collimated synchrotron threads linking the tails of ESO 137-006 from MeerKAT view.
Our simulation shows the rich magnetic threads.
However, the radio emission from this threads is not identified in the radio maps, because CRe in this region are already cooled and those CRe are not accelerated efficiently due to lower dissipation energies in this region.
An additional mechanism is needed to produce radio threads. 
The detailed discussion of the radio threads is described in the section \ref{sec:diffusion}.

%%------------------%%
%%-----------------%%
\subsection{Non-thermal X-ray emission}
\begin{figure*}
  \begin{minipage}{0.5\hsize}
    \includegraphics[width=1.0\textwidth]{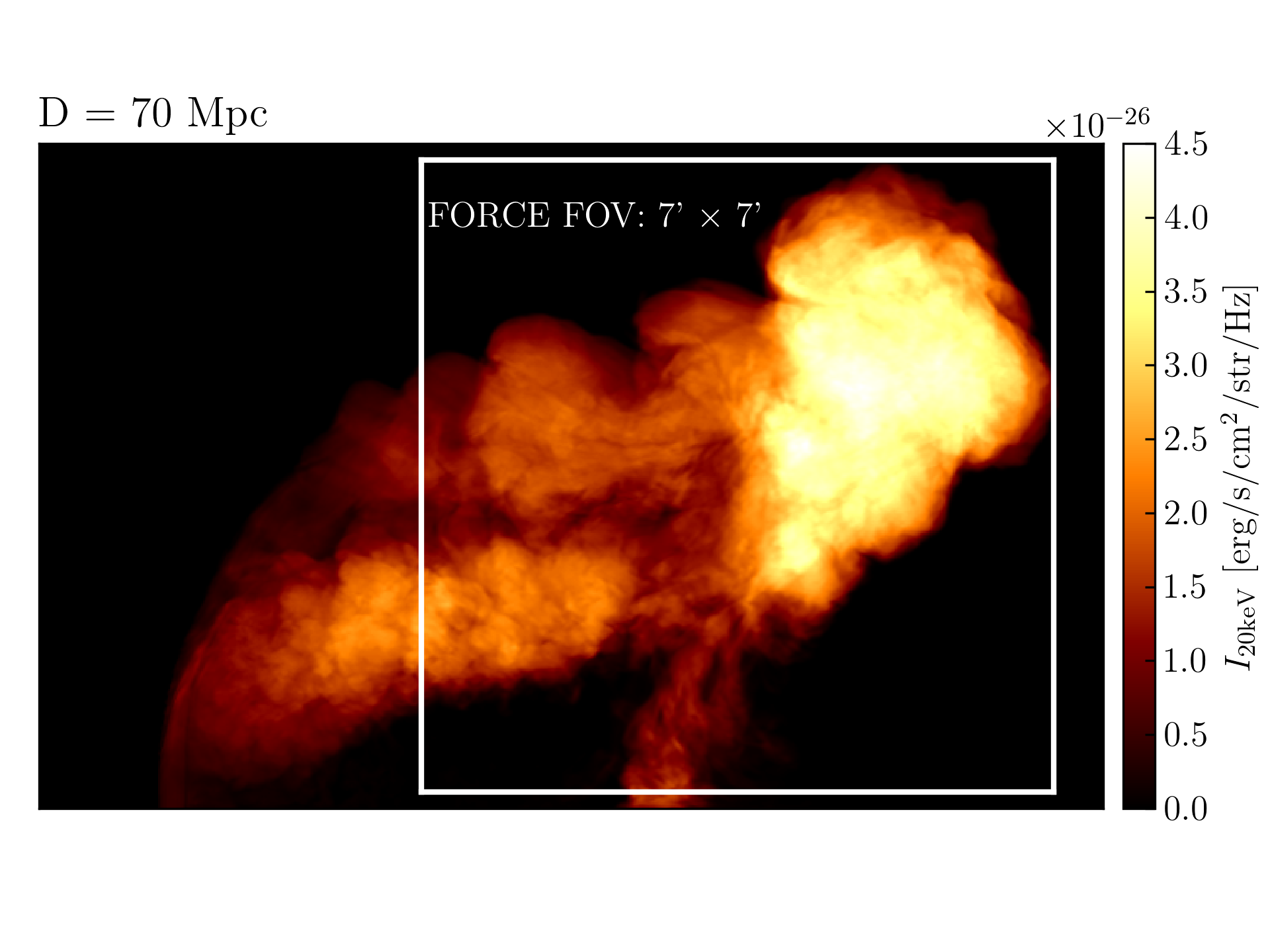}
  \end{minipage}
  \begin{minipage}{0.5\hsize}
    \includegraphics[width=1.0\textwidth]{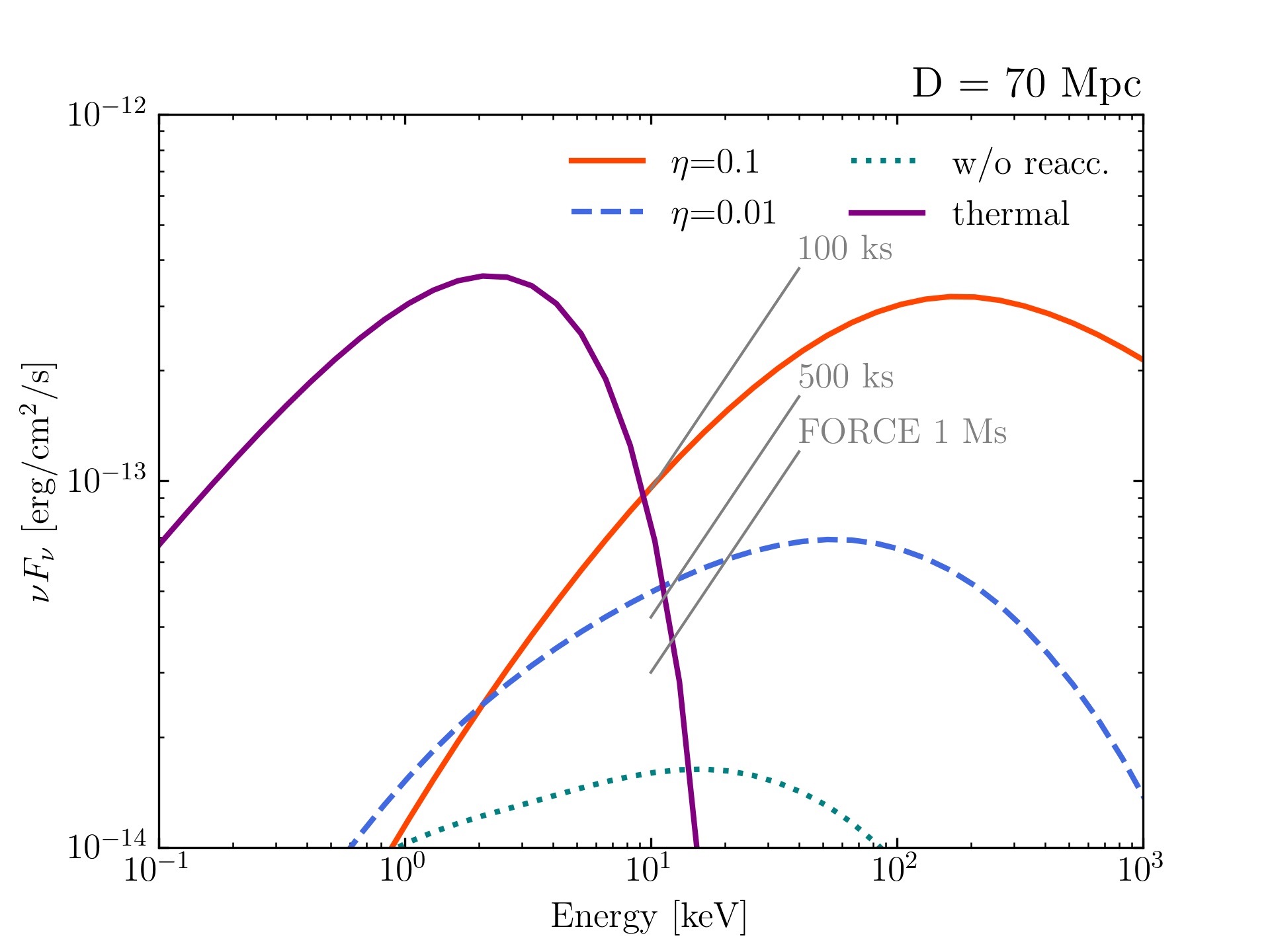}
  \end{minipage}
  \caption{ 
  \textbf{left}: 20 keV X-ray map for $\eta = 0.1$. The square corresponds to the FOV of FORCE at a distance of 70 Mpc. 
  \textbf{right}: X-ray spectra within the FOV of FORCE. The dashed blue line and solid red line show non-thermal X-ray spectra of inverse Compton scattering for $\eta = 0.01$ and $0.1$, respectively. The purple solid line shows the thermal X-ray emission from 5 keV ICM. The gray lines illustrate the target sensitivity of FORCE with 100 ks, 500 ks, and 1 Ms of observation time \citep{2018SPIE10699E..2DN}.
  }
  \label{fig:IC}
\end{figure*}

Figure \ref{fig:IC} (left) shows the 20 keV X-ray map for $\eta = 0.1$.
The typical Lorentz factor of 20 keV X-ray emitting electrons by inverse Compton scattering is $\gamma_{\rm IC} \sim 6000$.
The hard X-ray is the brightest in the termination of the radio tail, where the accelerated CRe are accumulated (see also the bottom right panel of Figure \ref{fig:ro_by}). 
This result is different from the radio morphology (see Figure \ref{fig:radiomap}), which is brightest at the bending point. 
% Also, one can see that relativistic electrons are stored inside the curve of jet, where the small-scale KH vortices well develop.

To discuss the detectability with the future X-ray mission, Focusing On Relativistic universe and Cosmic Evolution (FORCE) \citep{2016SPIE.9905E..1OM}, we derive X-ray spectra integrated within the field of view (FOV) of FORCE (Figure \ref{fig:IC}).
FORCE has $\sim$ 15" angular resolution, and thus has high sensitivity of $3 \times 10^{-15}$ ${\rm erg~cm^{-2}~s^{-1}~keV^{-1}}$ for 1 Ms observations in the range of 10 to 40 keV for point-like sources \citep{2018SPIE10699E..2DN}.
As shown in Figure \ref{fig:IC} (right), hard X-ray emissions from radio tail can be detected when electrons are reaccelerated efficiently.
% i.e., when more than 10 \% energy of jet kinetic energy is converted into CR particle energy.
Since the FORCE target sensitivity for diffuse sources is lower than that for point-like sources, it might be hard to detect hard X-ray emission for the case of inefficient reacceleration ($\eta < 0.01$).
% Therefore, the future X-ray observation may constrain the efficiency of CR reacceleration.

%%------------------%%
%%-----------------%%

\subsection{Gamma-ray and neutrino emissions}
\begin{figure}[t]
    \centering
    \includegraphics[width=0.5\textwidth]{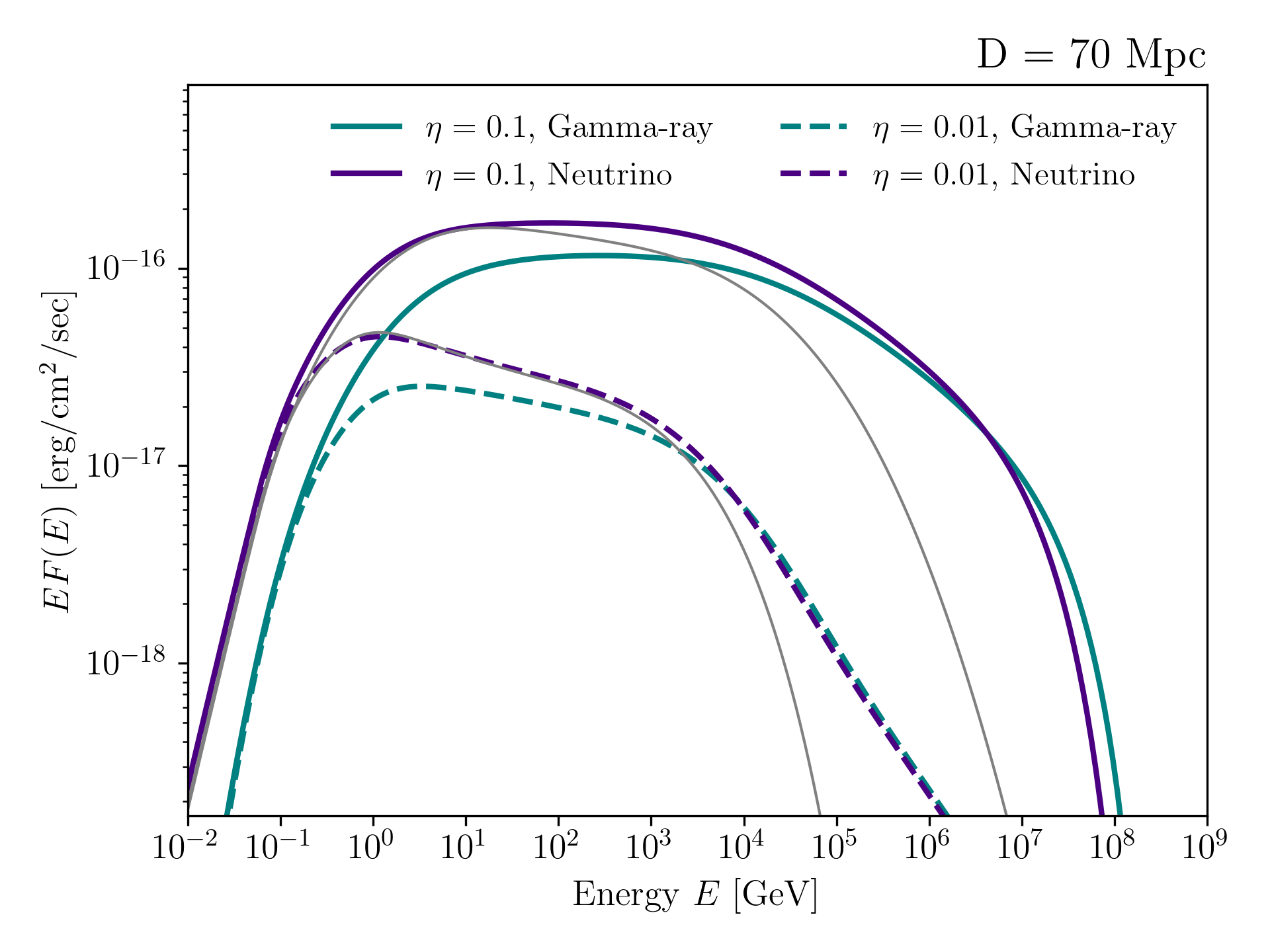}
    \caption{
    $\pi^0$ gamma (green) and neutrino (purple) spectra for $\eta = 0.01$ (dashed) and 0.1 (solid).
    The gray lines show the neutrino spectra of the one-zone model with $\tau_{\rm acc} =$ 100 (upper) and 800 (lower) Myr. }
    \label{fig:gamma_neutrino}
\end{figure}

We show gamma and neutrino spectra for $\eta = 0.01$ and $0.1$ in Figure \ref{fig:gamma_neutrino}.
The fluxes are calculated as a point source at $D=70$ Mpc.
% due to the lack of resolutions in gamma-ray and neutrino observations.
Note that, for simplicity, the extragalactic background light absorption, which is significant above 100 GeV, is ignored. 
These fluxes are more than three orders of magnitude below the upper limits of the Fermi-LAT and the IceCube.

Comparing with the spectra for the one-zone model (see detail in section \ref{subsec:CR energy}), our MHD models show harder spectra, especially for $\eta = 0.1$.
Note that the normalization of the one-zone model is adjusted to visualize the differences clearly in Figure \ref{fig:gamma_neutrino}.
Because turbulence develops around the contact discontinuity, CR particles are efficiently accelerated in these regions.
The gas density is also higher due to mixing of the ICM and the jet gas.
Therefore, contributions from such regions are dominant in the final spectra.

Only one candidate gamma-ray source has been found for head-tail galaxies.
\citet{2010A&A...519L...6N} reported that very high energy gamma-ray photons come from the head-tail galaxy IC 310.
However, its core has a blazer-like radio structure \citep{2012A&A...538L...1K}, i.e., the jet orients close to the line of sight.
% Such low-viewing angle is also independently supported from the morphological features of high-resolution VLA images \citep{2020MNRAS.499.5791G}.
Thus, high-energy TeV photons most likely originate from the core region.
Our result is compatible with this interpretation.

\section{spatial diffusion of electrons in the magnetic threads} \label{sec:diffusion}
% \subsection{The spatial diffusion of CRe in the magnetic threads} \label{sec:diffusion}
% In this section, we discuss about the electron spatial transportation process in the magnetic threads.
Our simulations show that head-tail galaxies have magnetic threads linking the two tails (see Figure \ref{fig:3D_thread}).
However, the radio intensity from these threads is lower than sensitivities of recent radio detectors \citep{2020A&A...636L...1R}.
In this section, we dicsuss the spatial diffusion effect, neglected in our simulations, on the radio brightness of the magnetic threads.
High-energy electrons penetrating into the threads would produce radio threads that can be identified with observations.
% Thus, the other physical processes, not considered in our simulations, may be required to transport the fresh CRe into the magnetic threads. 
% It is note that efficient reacceleration of CRe cannot expect to work since our results suggest that the dissipation rate in the threads is low.

First, we calculate the characteristic scales of magnetic field in the parallel and perpendicular in the magnetic threads and tails as follows \citep{2004ApJ...612..276S,2011ApJ...739...82B}
\begin{equation}
    l_{\parallel} = \sqrt{\frac{\langle\bm{B}^4\rangle}{\langle|\bm{B} \cdot \nabla \bm{B}|^2\rangle}},
\end{equation}
\begin{equation}
    l_{\perp} = \sqrt{\frac{\langle\bm{B}^4\rangle}{\langle|\bm{B} \times \bm{J}|^2\rangle}},
\end{equation}
where $\bm{J} = \nabla \times \bm{B}$, and $\langle\rangle$ denotes a volume average.
% The regions are set to $x_{t} \in [108,126]$ kpc, $y_{t} \in [-52,23]$ kpc, and $z_{t} \in [-4,4]$ kpc for the magnetic threads and $x_{l} \in [108,183]$ kpc, $y_{l} \in [88,107]$ kpc, and $ z_{l} \in [-4,4]$ kpc for the tail (see the rectangles in Figure \ref{fig:ro_by}) , respectively.
The spatial regions for the average are shown by the rectangles in Figure \ref{fig:ro_by}.
The tail and threads have similar values of $l_{\perp}$, 1.35 kpc and 1.71 kpc, respectively.
In contrast to this, we can see the magnetic structures of the threads are more anisotropic than that of the tail.
The values of $l_{\parallel}$ are 3.99 kpc for tails and 9.88 kpc for threads.
Here, we note that a highly anisotropy with $l_{\parallel}/l_{\perp} > 10$ can be seen when we cut out a small region focusing on the large magnetic threads.

% We now consider spatial diffusion of CRe via pitch angle scattering interact with the magnetic field.
First, we discuss the spatial diffusion of CRe. 
Based on the result of the previous paragraph, we assume that CRe can diffuse only along the flux-tube.
Typical value of magnetic field in the threads is 5 $\mu$G, and hence the cooling time for 10 GeV electrons is $\tau_{\rm cool} \sim 170$ Myr.
The diffusion length can be written as $l_{\rm diff} \sim 2 \sqrt{D_{\parallel}t}$ in a time $t$, where $D_{\parallel}$ is the spatial diffusion coefficient.
Thus, the spatial diffusion coefficient along the flux tube required for transportation of CRe to shine the threads, whose half length is about 70 kpc, within the cooling time is $\sim 2.1 \times 10^{30}$ $\rm cm^2/s$.
% Meanwhile, recent cosmological zoom-in CR MHD simulations reproduce observed $\gamma$-ray emissivities with $D \sim 3-30 \times 10^{29}$ $\rm cm^2/s$ \citep{2020MNRAS.492.3465H}.

Considering the spatial diffusion via the gyroresonant interaction, the spatial diffusion coefficient can be written as \citep{1969ApJ...156..445K}
\begin{equation}
\label{eq:diffusion}
D_{\parallel} = \frac{1}{3} c r_{\rm g}  \qty( \frac{B^2_{\parallel}}{k_{\rm g} W_{B_{\perp}}(k_{\rm g})} ),    
\end{equation}
where $r_{\rm g},~k_{\rm g}$ and $W_{B_{\perp}}$ are the Larmor radius of the CRe, $k_{\rm g} = 2\pi r^{-1}_{\rm g}$, and the turbulent spectrum of the perpendicular magnetic field, respectively. 
We compute the volume average parallel magnetic field in the region of the thread as 
\begin{equation}
\bar{B_{\parallel}} = \frac{\int B_y N_{\rm e, 100MeV} \dd{V}}{\int N_{\rm e, 100MeV} \dd{V}} = 4.8~\mu{\rm G},  
\end{equation}
and the average perpendicular magnetic field as 
\begin{equation}
\label{eq:Bperpav}
\bar{B_{\perp}}(y) = \frac{\int \int  B_{\perp} N_{\rm e, 100MeV} \dd{x}\dd{z}}{\int \int N_{\rm e, 100MeV} \dd{x}\dd{z}},
\end{equation}
where $B_{\perp} = \sqrt{B_x^2+B_z^2}$.
Then, we calculate the turbulence spectrum by using the Fourier transform of $\bar{B_{\perp}}(y)$.
We find that the index of the turbulence spectrum is close to the Kolmogorov scaling, $W_{B_{\perp}} \propto k^{-5/3}$.
Since the Larmor radius of 10 GeV electrons is much smaller than the MHD grid size, the resonant occurs at the sub-grid scale.
Therefore, we extrapolate the turbulence spectrum extrapolating the Kolmogorov scaling to estimate $D_{\parallel}$ for 10 GeV electrons.
As a result, we obtain $D_{\parallel} \sim 10^{32}~{\rm cm^2/s}$ from equation \eqref{eq:diffusion}.
This value is two orders of magnitude larger than that required for the transportation of CRe.
Meanwhile, this discussion is optimistic in the sense of assuming that the flux tube completely lies in the y-direction, i.e., the flux tube is not tangled.
% This is higher than the standard value of the spatial diffusion coefficient in the Galaxy $D = 10^{28}~(E/10 ~{\rm GeV})^{1/2} (B/ 3~{\rm \mu G})^{-1/2}$ $\rm cm^2/s$ \citep[e.g.,][]{1990acr..book.....B,2010ApJ...712L.153F}.
The average procedure with equation (\ref{eq:Bperpav}) may also lead to an underestimated amplitude of the local turbulence. 

% CR streaming 
% Second, the CR streaming may be another possible process \citep{1969ApJ...156..445K}.
In addition to the spatial diffusion, "CR streaming" is often invoked as another transport mechanism of CR particles \citep{1969ApJ...156..445K}. 
The CR particles can stream along the magnetic field line with Alfv\'{e}n waves, and it is generally known that the streaming velocity $v_{\rm st}$ is limit to Alfv\'{e}n velocity $v_{\rm A}(=B/\sqrt{4\pi \rho})$.
% But, in general, the streaming velocity remains (deeply) uncertain.
Optimistically, we adopt $B = 5~\mu{\rm G}, \rho=1.0\times10^{-3} m_{\rm p}$ and $v_{\rm st} = v_{\rm A} \sim 335$ km/s.
So that the length that CRe can stream along the flux tube in $\tau_{\rm cool}$ is $l_{\rm st}= v_{\rm A} \tau_{\rm cool} \sim 67$ kpc, which is comparable to the length of threads.
This simple estimate implies that the CR streaming could also be the important process for the formation of synchrotron threads.

\section{Summary and discussion} \label{sec:summary}
We have performed the 3D CR MHD simulations of a head-tail galaxy, focusing on reacceleration of CR particles and non-thermal emissions.
Because the scale at which CR particles interact with turbulence is much 
smaller than the resolution of our simulations, we estimate the efficiency of reacceleration with a sub-grid recipe to bride this scale gap.
% the timescale of reacceleration should be implemented through a sub-grid recipe to bride this scale gap.
In this paper, we adopt the hard-sphere type acceleration model with a parameter $\eta$, that adjusts the energy conversion efficiency from dissipation energy to the CR particle energy.
The main results of this study are summarized as follows:
\begin{itemize}
    \item Depending on the values of $\eta$, 3 - 30 \% of the jet kinetic energy is converted into CR particles energy at $t=316.3$ Myr. Our global simulations show significantly different CR spectra from those in simple one-zone models. 
    % \item Integrated radio spectrum of the acceleration efficient model ten to be harder than that of the acceleration inefficient model around GHz. The bump of radio flux and the flattening of the spectral index at the tip of jets can be seen for reacceleration models because the CRe, which are accelerated for a long time, are deposited there.
    \item In the presence of reacceleartion, the radio flux and spectral index do not decrease along with the tails. Those behaviors are consistent with some head-tail galaxies, and therefore reacceleration is essential. For $\eta = 0.1$, the spectral index for hundreds of MHz frequency ranges is harder than that for $\eta = 0.01$ and the radio observations.
    \item Inverse Compton X-ray emission from head-tail galaxies in the Perseus Cluster can be observed by the future X-ray observatory (FORCE). On the other hand, hadronic gamma-ray and neutrino emissions are too dim to be detected with the current instruments. 
    \item  Thin magnetic field threads connecting the two tails are identified in our simulations. The origin of these threads is backflow at early phase. The backflowing materials are simply advected by the wind to form the threads. 
    \item An efficient transport mechanism of CRe is needed to explain the observed radio threads. Considering the spatial diffusion process of CRe along the threads, $D_{\parallel} \sim 2.1 \times 10^{30}$ $\rm cm^2/s$ is required.
\end{itemize}

%結果がNeとNpの入れ方にも依存することについて
% There is one caveat that the reacceleration timescale is not only dependent on the dissipation energy, but also the total CR energy from our sub-grid model.
% Injection parameters of CRe and CRp therefore influence for our results.
% From radio and X-ray observation of radio jets, the energy density of the magnetic field and CRe can be estimated, and these energy are comparable, though the energy density of CRe is slightly higher than that of magnetic field \citep{2002ApJ...581..948H, 2004ApJ...612..729H}.
% However, it is difficult to constrain the ratio of energy in CRp to CRe from observation while a significant contribution of non-radiating particles (CRp and/or thermal particles) to inflate the radio lobe of the FR-I jets is needed \citep{2000MNRAS.319..562H,2014MNRAS.438.3310C}.
% Therefore, in this work, we simply choice that $u_{\rm e} \approx u_{\rm p}$ at the jet injection point.
% But, we comment that a large $\eta$ value is needed to obtain same acceleration history of CRe if $u_{\rm p} \gg u_{\rm e}$.

This study is our first step in constructing a realistic model to implement the reacceleration process with MHD simulations, and there are several important limitations in our current models.
% Feedback process
First, our code do not account for the dynamical back-reaction from CR particles.
The CR energy density is comparable to the thermal energy density for $\eta = 0.1$.
Therefore, the CR pressure may be dynamically significant.

% Spatial diffusion process
Second, the spatial diffusion and streaming would affect the radio emission map.
As discussed in section \ref{sec:diffusion}, it would be important for the formation of synchrotron threads. 
Meanwhile, in the tail regions, the magnetic field is highly tangled so that the spatial diffusion may be suppressed in such disturbed regions.
This picture has been supported by early MHD simulations of radio jets \citep{2018MNRAS.481.2878E}.
% Confinement of CR particles due to tangled magnetic field can be seen in MHD jets \citep{2018MNRAS.481.2878E}.

% DSA mechanisim
Additionally, subsequent particle injection has been neglected.
% We justify this simplification because the Mach number of the jet is lower than a critical Mach number for DSA. 
\citet{2019ApJ...884...12O} showed that weak shocks in the tail region is too weak to inject particles.
However, the shock acceleration can be dominant process in high power jets.
As future extension of this study we should implement those processes in a self-consistently manner \citep[e.g.,][]{2020MNRAS.491..993G,2021ApJS..253...18O,2022arXiv220705087B}.

% \noindent author year, title, version, publisher, prefix:identifier\\

\acknowledgments

We thank the anonymous referee for the useful comments that greatly improved the presentation of the paper. 
We are grateful Reinout van Weern, Kyohei Kawaguchi, Tomoya Kinugawa, Tomohisa Kawashima, and Hiroki Akamatsu to for fruitful discussion.
This work was supported by JSPS KAKENHI Grant Numbers JP22K14032 (T.O.), 22K03684 23H04899 (K.A.), 19K03916 (M.M.), 20J13339, and 22K20386 (H.S.).
This work is supported by the joint research program of the Institute for Cosmic Ray Research (ICRR).
K.N. is supported by FoPM, WINGS Program, the University of Tokyo.
Our numerical computations were carried out on the Cray XC50 at the Center for Computational Astrophysics of the National Astronomical Observatory of Japan.
The computation was carried out using the computer resource by Research Institute for Information Technology, Kyushu University.
This work was also supported in part by MEXT as a priority issue (Elucidation of the fundamental laws and evolution of the universe) to be tackled by using post-K Computer and JICFuS and by MEXT as “Program for Promoting Researches on the Supercomputer Fugaku” (Toward a unified view of the universe: from large scale structures to planets).
% \appendix

% \section{Appendix information}

% Appendices can be broken into separate sections just like in the main text.

\bibliography{refapj}

\end{document}